\documentclass[aps,prb,twocolumn,superscriptaddress,showpacs]{revtex4}
\usepackage{multirow}
\usepackage{graphicx}
\usepackage{longtable}
\usepackage[utf8]{inputenc}
\usepackage{epstopdf}
\usepackage{longtable}
\usepackage{color}
\newcommand{\minitab}[2][l]{\begin{tabular}{#1}#2\end{tabular}}
\newcommand{\kreis}[1]{\unitlength1ex\begin{picture}(2.5,2.5)%
\put(0.75,0.75){\circle{2.5}}\put(0.75,0.75){\makebox(0,0){#1}}\end{picture}}

\begin{document}

\title{Unexpected trend of exchange interactions in Fe clusters on Rh(111) and Ru(0001)}

\author{F. Otte}
\email{otte@theo-physik.uni-kiel.de}
\affiliation{Institut f\"ur Theoretische Physik und Astrophysik,
Christian-Albrechts-Universit\"at zu Kiel, D-24098 Kiel, Germany}

\author{P. Ferriani}
\affiliation{Institut f\"ur Theoretische Physik und Astrophysik,
Christian-Albrechts-Universit\"at zu Kiel, D-24098 Kiel, Germany}

\author{S. Heinze}
\affiliation{Institut f\"ur Theoretische Physik und Astrophysik,
Christian-Albrechts-Universit\"at zu Kiel, D-24098 Kiel, Germany}

\date{\today}

\begin{abstract}
We use first-principles calculations based on density functional theory
to investigate the magnetic exchange interaction of Fe clusters on
Rh(111) and Ru(0001).
We consider dimers, trimers, tetramers, and pentamers of different shape in fcc and hcp stacking
as well as infinite atomic and biatomic chains.
From the dimer calculations we extract the exchange interaction as a function of adatom distance
by mapping total energies to a Heisenberg model. The nearest-neighbor (NN) exchange constant is
about one order of magnitude smaller than reported for other substrates
due to the strong hybridization between the Fe atoms and the partly filled $4d$-band of the
surface.
We also find a transition from a ferromagnetic NN exchange interaction for Fe dimers on Rh(111)
to an antiferromagnetic one on Ru(0001).
The distance-dependent
exchange coupling displays a RKKY-like oscillatory behavior which is nearly inverted for Fe dimers on the Rh(111)
surface compared to those on Ru(0001).
Unexpectedly, for Fe clusters beyond dimers, a complex trend of the magnetic ground state
is observed which alternates between ferro- and antiferromagnetic configurations depending on cluster size and shape.
In view of the exchange constants obtained for dimers, it is surprising that
on both surfaces small compact clusters are ferromagnetic while open structures such as
linear trimers or tetramers become antiferromagnetic. We demonstrate that both vertical and lateral structural relaxations
of the clusters are crucial in order to understand this
unexpected
trend of magnetic order and connected to the competition of direct ferromagnetic exchange among Fe atoms in the cluster and
the hybridization with the substrate.
\end{abstract}


\maketitle

\section{Introduction}
The discovery of the giant magnetoresistance by Fert and Gr\"unberg~\cite{BBF1988,BGS1989} 
initiated the field of spintronics which aims at utilizing the electron spin degree of freedom for storage and
transportation of information.
Today, the possibility opened by scanning tunneling microscopy (STM) to manipulate systems with atomic
precision \citep{IBM} and to detect the magnetic state of single atoms \citep{Hirj,Meier,Zhou,Serrate2010,Loth2010a}
using spin-polarized STM \cite{Bod2003,Wiesendanger2009} and inelastic scanning tunneling spectroscopy
\cite{Heinrich2004,Loth2010b}
allows the exploration of spintronic concepts at the atomic level. Some recent progress in this direction
have been the demonstration of the spin-valve effect~\cite{Ziegler2011} and
tunneling anisotropic magnetoresistance~\cite{Neel2013} at the single-atom limit,
the control of the atom spin state by electric currents~\cite{Loth2010c}, the creation of a spin
logic gate consisting of only a few atoms \citep{Khaj}, the atomic engineering of nanomagnets \citep{Khaj2},
and the demonstration of storing information on antiferromagnetic clusters \citep{Loth13}.
In order to interpret such experiments and to develop novel nanomagnets with tailored properties
there is a need for a microscopic understanding of the exchange interactions in magnetic
nanostructures at surfaces. First-principles electronic structure calculations have therefore
become an indispensable tool in this field.

Recently, such studies based on density functional theory (DFT)
showed that the nearest-neighbor exchange interaction in Fe monolayer (ML) films can be systematically tuned
from ferro- to antiferromagnetic
by changing the $d$-band filling of a non-magnetic transition-metal (TM) substrate \cite{Fer2007,Hardrat}.
A transition occurs e.g.~between the Fe ML on Ru(0001), possessing an antiferromagnetic (AFM) exchange coupling, and the Fe ML on Rh(111),
which has a ferromagnetic (FM) exchange. Since the energy difference between the FM and the AFM state is small for these
two systems, exchange beyond nearest-neighbors and higher-order interactions
become important and may lead to complex magnetic ground states. For Fe/Ru(0001) a 120$^{\circ}{}$ N\'eel state has been theoretically
predicted and for Fe/Rh(111) a double row-wise antiferromagnetic structure also denoted as the $uudd$-state has been
proposed~\citep{Hardrat,alzubiPaper}. In particular,
the $uudd$-state is very intriguing since it cannot be understood based on mapping
the total energies from a DFT calculation to a Heisenberg model
but is closely linked to the large induced magnetic moments in the Rh substrate\citep{alzubiPaper}.
Up to now, however, these predictions still lack experimental verification.

Moving to transition-metal clusters of a few atoms up to small islands additional effects originating from the cluster size, shape,
and geometry come into play.
An example is the linear behavior of the magnetic moment on the coordination which has
been found for Fe clusters on nonmagnetic surfaces based on first-principles calculations \citep{Mavro}.
In most of the theoretical studies so far
Fe clusters have been regarded on metal surfaces with a filled $d$-shell
such as Cu(111)
\citep{Mavro,FeMnCrCu111} or Pd(111)\citep{FePd} where nearest-neighbor (NN) exchange coupling is strongly FM. However,
the magnitude depends
sensitively on cluster geometry, shape, and size\citep{Mavro}. The same holds for Co clusters which have been investigated on Pd(111)
and Au(111) \citep{Sipr2007}. Namely, the NN
exchange coupling tends to decrease with increasing cluster size
and corner-shaped trimers show stronger FM exchange than linear ones.
A recent DFT study included Fe clusters on the Ir(111) surface \cite{Ir111} and reported the dependence of the
exchange interaction on the cluster size and geometry.
Unfortunately, structural relaxations were not included although
they are quite important in such systems as we demonstrate here.
Very recently, it has been predicted based on DFT calculations that small Co clusters on W(110) possess antiferromagnetic or
ferrimagnetic ground states~\cite{Lukashev2013} illustrating the key impact of the substrate.
The antiferromagnetic coupling in Mn clusters leads to a complex trend of noncollinear or AFM states depending on
the geometry\citep{FeMnCrCu111,MnPt}. Experimentally, the RKKY-like oscillation of the exchange interaction between
$3d$-TM adatoms on metal surfaces has been measured applying
STM based on the Kondo effect~\cite{Wahl} and single-atom magnetization curves~\cite{Meier,Zhou}.

Here, we apply first-principles electronic structure calculations based on DFT to demonstrate that the modification of the exchange
coupling for Fe clusters due to
the hybridization with the Ru(0001) or Rh(111) surface leads to a complex evolution of the magnetic ground state depending on cluster
size, shape, and geometry. 
Unexpectedly, many of the Fe clusters display a compensated antiferromagnetic or ferrimagnetic state.
We take
full structural relaxations of the clusters into account and find that they are crucial for the
hybridization of Fe atoms within the cluster and with the substrate
and thus for the magnetic ground state. From dimer calculations we obtain
the exchange constants as a function of Fe adatom separation. 
Due to the hybridization with the partly filled $4d$-band of the Rh or Ru substrate, the 
nearest-neighbor (NN) exchange interaction is 
an order of magnitude than reported for other substrates.
The NN exchange interaction favors ferromagnetic coupling on Rh(111), while it is
antiferromagnetic on Ru(0001). The distance dependence of the exchange interaction displays a RKKY-like oscillation 
and the trend is nearly inverted when comparing Fe dimers on Rh(111) with those on Ru(0001).

We find on both substrates that small compact clusters such as trimers and tetramers possess a FM ground state,
while more open geometries such as linear and corner-shaped trimers and tetramers lead to an antiferromagnetic ground state.
The FM state
of compact Fe clusters on Ru(0001) is surprising in view of the antiferromagnetic NN exchange coupling found for the
dimers and can be explained based on the lateral structural relaxation and direct exchange in the cluster. In contrast,
Fe pentamers on Ru(0001) display an antiferromagnetic ground state since the extra atom added to a tetramer alters
the structural relaxation within the cluster and weakens the direct FM exchange between Fe atoms.

While the AFM order in corner-shaped trimers on Rh(111) results from the interplay of nearest and next-nearest exchange
interaction, the weakening of the direct FM exchange between Fe atoms explains the antiferromagnetic state in the linear trimer.
We further demonstrate that infinite atomic, biatomic, and triatomic Fe chains on Rh(111)
already have a tendency to favor the $uudd$-state proposed as the ground state of the full monolayer.

The paper is structured as follows. After introducing the computational method and details in section \ref{sec:method},
we start in section \ref{sec:adatoms} with a discussion of the properties of single Fe adatoms and dimers
on Rh(111) and Ru(0001) with varying distance between the adatoms.
Subsequently, in sections \ref{sec:clusters} and \ref{sec:open_tetramers},
clusters with up to three and four atoms with different geometries are analyzed.
The important role of structural relaxations on the exchange interactions in these
systems is stressed and the interplay of direct exchange between Fe atoms and the hybridization with the substrate
is discussed. As a first step to study the transition to the monolayer we consider Fe pentamers on Ru(0001) in
section \ref{sec:pentamers}
and infinite atomic and biatomic Fe chains on Rh(111) in section \ref{sec:chains}.
A summary and conclusions are given in the final section.

\section{Computational Details}
\label{sec:method}
Fe clusters on Rh(111) and Ru(0001) have been studied based on density functional theory
calculations in the generalized gradient approximation (GGA) to the
exchange-correlation functional\citep{PBE}, using the projecter-augmented-wave method as implemented in the Vienna Ab-Initio Simulation Package (VASP)~\cite{VASP1,VASP2,VASP3,VASP4,VASPPAW}.
All calculations have been performed in the scalar-relativistic approximation, i.e.
neglecting the effect of spin-orbit coupling.
To model the Fe clusters we have used the $p(4\times4){}$ surface unit cell and eight layers of substrate
to model the Rh(111) or Ru(0001) surface. The adatoms as well as the two upmost surface layers have been
structurally relaxed until the forces were smaller than 0.005 eV/\AA. A ($5\times 5 \times 1$) $\Gamma$-centered
k-point mesh has been used. The experimental lattice constant of 3.8034~{\AA} for Rh and lattice parameters of
2.7059~{\AA} and 4.2815~{\AA} for Ru have been chosen
which differ by only about 1\% from the theoretical values.
The energy cutoff parameter for the plane wave expansion
was 390 eV and a Gaussian smearing of $\sigma=0.07{}$ eV has been applied.

\begin{figure}
\includegraphics[width=0.9\linewidth]{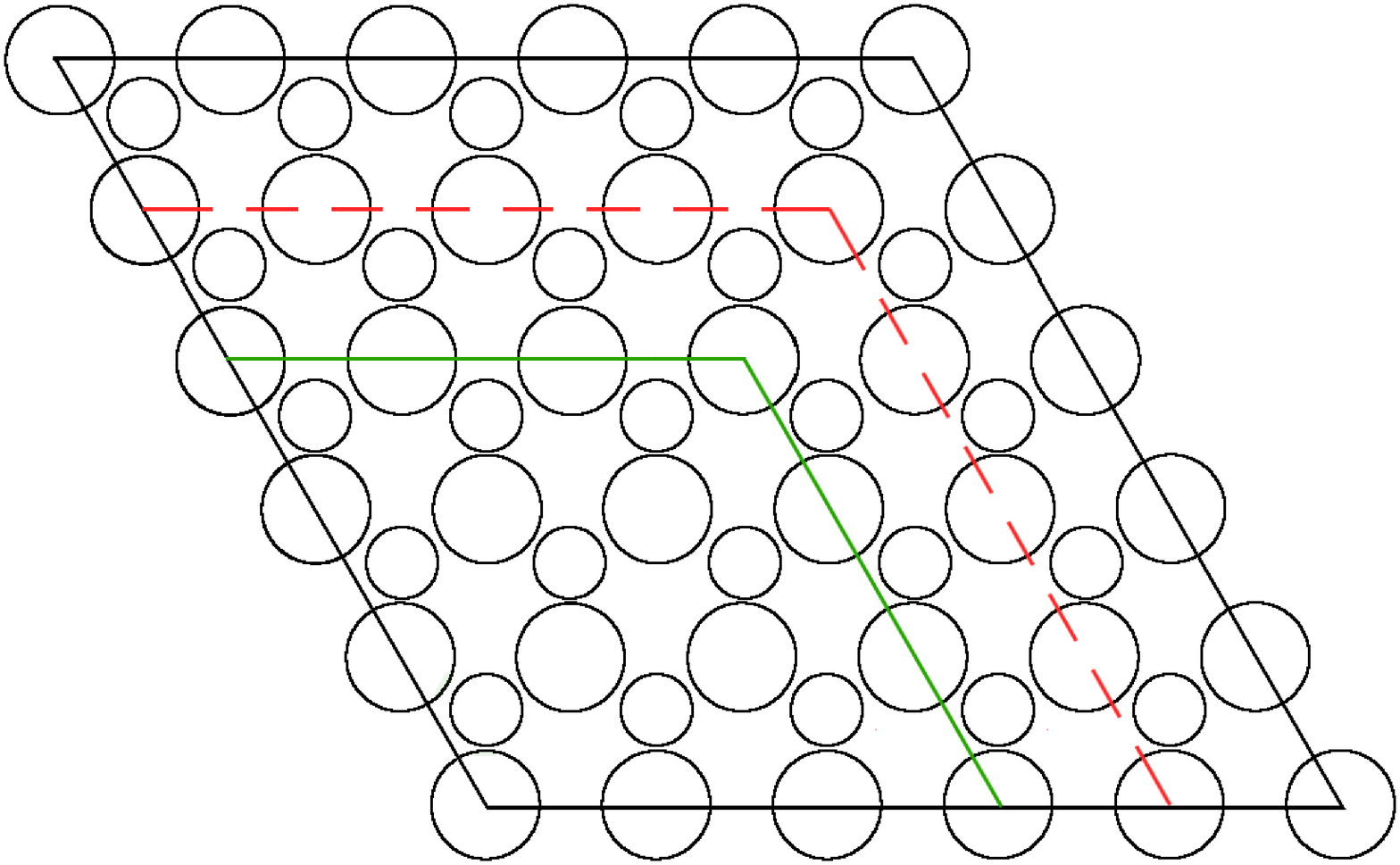}
\caption[fig1]{(Color online) Sketches of the $p(3\times3){}$ (green), $p(4\times4){}$ (red and dashed), and $p(5\times5){}$ (black) unit cell of the (111) or (0001) surface used in the calculations. Large circles denote the surface atom layer and small circles the subsurface
layer.}
\label{fig:unit_cell}
\end{figure}

An important aspect of our approach to calculate the exchange constants is the possible
interaction of the clusters with those in adjacent cells
due to the two-dimensional (2D) periodic boundary conditions. In order to estimate the influence of atoms in
neighboring cells we have performed test calculations in the $p(3\times3)$, $p(4\times4)$, and $p(5\times5){}$ unit
cells, which are depicted in Fig.~\ref{fig:unit_cell}.
These tests are mentioned in the text during the discussion of the respective systems.
We found that the $p(4\times4){}$ unit cell size is sufficiently large to avoid spurious interaction effects
for compact clusters and to determine their magnetic ground state.
However, if the distance between the adatoms within the unit cell is large
as for dimers with large separation,
the influence of atoms in the adjacent unit cells also becomes important. We have taken such interactions into account
when determining the exchange constants. Details are given in section \ref{sec:adatoms}.

\section{Adatoms and dimers}

\label{sec:adatoms}

\begin{table}
\caption{\label{tab1}Structural relaxations of Fe adatoms and dimers in ferro- ($\uparrow \uparrow$) and antiferromagnetic ($\uparrow\downarrow$) coupling on Rh(111) and Ru(0001). The relative vertical relaxation is defined as
$\Delta_{12}=(z_{12}-z_{0})/z_0$, where $z_{12}$ is the vertical distance of the Fe adatom to the nearest-neighbor
surface atoms and $z_0$ is the unrelaxed value for the Rh(111) or Ru(0001) surface. A negative sign denotes a
relaxation of the atoms towards the surface.
For the mixed dimers relaxations for atoms in hcp and fcc positions are shown individually.
$d_{\rm{Fe-Fe}}$ denotes the distance between the Fe atoms in the dimer after relaxation. }
\begin{ruledtabular}
\begin{tabular}{lccccccc}


Fe on&  &  \multicolumn{4}{c}{Rh(111)\hspace{1cm}}		 &  \multicolumn{2}{c}{Ru(0001)\hspace{1cm}}   \\\hline
    & & \multicolumn{2}{c}{$\Delta_{12}{}$ (\%)}& $d_{\rm{Fe-Fe}}{}$ (\AA)&  \multicolumn{2}{c}{$\Delta_{12}{}$ (\%)}& $d_{\rm{Fe-Fe}}{}$ (\AA)    \\\hline

  adatom (hcp)&&  \multicolumn{2}{c}{$-21$}&  & \multicolumn{2}{c}{$-17$}& \\\hline
  dimers (hcp) &&  \multicolumn{2}{c}{}&  & \multicolumn{2}{c}{}& \\\hline

  \multirow{2}{*}{2.7~{\AA}}& $\uparrow \uparrow$ &  \multicolumn{2}{c}{$-18$}&2.63 & \multicolumn{2}{c}{$-15$} &2.67  \\
                      & $\uparrow \downarrow$  & \multicolumn{2}{c}{$-19$} &2.70& \multicolumn{2}{c}{$-15$} &2.72 \\ \hline

  \multirow{2}{*}{4.7~{\AA}}& $\uparrow \uparrow$ & \multicolumn{2}{c}{$-19$} &4.68 & \multicolumn{2}{c}{$-16$}&4.69   \\
                      & $\uparrow \downarrow$  & \multicolumn{2}{c}{$-19$} &4.68 & \multicolumn{2}{c}{$-16$}& 4.70 \\\hline

  \multirow{2}{*}{5.4~{\AA}}& $\uparrow \uparrow$  & \multicolumn{2}{c}{$-21$} &5.40& \multicolumn{2}{c}{$-17$} & 5.41  \\
                      & $\uparrow \downarrow$ & \multicolumn{2}{c}{$-21$}&5.40& \multicolumn{2}{c}{$-17$} & 5.41   \\\hline

\\\hline

  adatom (fcc)&&                               \multicolumn{2}{c}{$-20$}&  & \multicolumn{2}{c}{$-14$} &   \\\hline
  dimers (fcc)&&  \multicolumn{2}{c}{}&  & \multicolumn{2}{c}{}& \\\hline
  \multirow{2}{*}{2.7~{\AA}} &$\uparrow \uparrow$ & \multicolumn{2}{c}{$-16$}&2.60 & \multicolumn{2}{c}{$-11$} &2.61  \\
                        & $\uparrow \downarrow$  & \multicolumn{2}{c}{$-18$}&2.72 & \multicolumn{2}{c}{$-12$} &2.69  \\\hline

  \multirow{2}{*}{4.7~{\AA}}& $\uparrow \uparrow$ & \multicolumn{2}{c}{$-18$} &4.68 & \multicolumn{2}{c}{$-13$}& 4.67   \\
                      & $\uparrow \downarrow$ & \multicolumn{2}{c}{$-18$}  &4.68 & \multicolumn{2}{c}{$-13$} & 4.70  \\\hline

  \multirow{2}{*}{5.4~{\AA}}& $\uparrow \uparrow$   & \multicolumn{2}{c}{$-20$}&5.40  & \multicolumn{2}{c}{$-14$} & 5.41 \\
                      & $\uparrow \downarrow$  & \multicolumn{2}{c}{$-20$}&5.40& \multicolumn{2}{c}{$-14$}&5.41 \\\hline

\\\hline

  mixed dimers & & hcp & fcc                                         &   &hcp  & fcc &                \\\hline
  \multirow{2}{*}{3.1~{\AA}}& $\uparrow \uparrow$& $-19$&$-18$&3.08& $-15$ & $-11$ &3.04   \\
                      & $\uparrow \downarrow$  & $-20$&$-19$&3.17 & $-16$& $-11$& 3.04  \\\hline

  \multirow{2}{*}{4.1~{\AA}}& $\uparrow \uparrow$ & $-21$&$-19$ &4.08& $-17$& $-13$ & 4.05  \\
                      & $\uparrow \downarrow$& $-21$&$-19$ &4.09 & $-17$& $-13$ & 4.09    \\\hline

  \multirow{2}{*}{5.6~{\AA}}& $\uparrow \uparrow$  & $-21$ & $-20$&5.63& $-17$&$-14$  & 5.64  \\
                      & $\uparrow \downarrow$  & $-21$&$-20$&5.63 & $-17$& $-14$ &5.63  \\\hline

  \multirow{2}{*}{6.2~{\AA}}& $\uparrow \uparrow$ & $-21$&$-20$&6.24 & $-17$& $-14$  & 6.25\\
                      & $\uparrow \downarrow$ & $-21$ &$-20$ &6.24& $-17$& $-14$ & 6.25 \\

\end{tabular}
\end{ruledtabular}
\end{table}

In this section we present the structural, electronic, and magnetic properties of Fe adatoms and dimers adsorbed on
Rh(111) and Ru(0001). We analyze in detail the effect of structural relaxations on the magnetic interactions in dimers.
We find that the nearest-neighbor exchange interaction in the Fe dimers is ferromagnetic on Rh(111) while it is antiferromagnetic
on Ru(0001). On both substrates the absolute value is by about one order of magnitude smaller than reported on other surfaces.
For exchange interactions beyond nearest neighbors we find an RKKY-like behavior which shows an nearly inverted trend
for Fe dimers on Rh(111) compared to those on Ru(0001).

\subsection{Adatoms - magnetic moments and relaxations}
As a reference system, we first consider Fe adatoms on Rh(111) and Ru(0001)
calculated in the $p(4\times4){}$ unit cell, i.e.~with a distance of  10.8~{\AA} between adatoms
in adjacent cells. The structural data after relaxation is presented in table \ref{tab1}.
For Fe adatoms on Rh(111), we find that adsorption of Fe atoms in the hcp site is
energetically more favorable by 87 meV/Fe-atom than in the fcc site.
We obtain a large relative vertical relaxation of the Fe adatoms towards the surface
of 21 \% for adsorption in the hcp position and 20 \% in the fcc position.
The small difference between the relaxations for hcp and fcc adsorption sites
can be attributed to the different position with respect to atoms in the second substrate layer.
The distance to the nearest substrate atom of the subsurface layer is larger in the fcc case
leading to smaller hybridization and in turn to smaller relaxations.
The magnetic moment of the Fe adatom amounts to 3.22~$\mu_B{}$ for both hcp and fcc stacking and the induced Rh moment
is 0.27~$\mu_B{}$ for hcp and 0.25~$\mu_B{}$ for fcc stacking.
This leads to a total magnetic moment per adatom of about 4~$\mu_B{}$.
These results compare well with those from Ref.~\onlinecite{Blonski}.

For Fe adatoms on Ru(0001), the vertical inward relaxation is smaller by 4 \% in hcp and 6 \% in fcc position than on Rh(111)
as seen in table \ref{tab1}. While the in-plane lattice constants of the substrates are very similar the hcp stacking in the Ru case
apparently plays a more important role.
In accordance the dependence of the relaxation on the adsorption site is more pronounced for adatoms on Ru(0001)
and the energy difference between the two adsorption sites is larger than on Rh and amounts to 198~meV/Fe-atom in favor of the hcp site.
The different relaxation also affects the magnetic moment of the adatom which is 2.97~$\mu_B{}$ in hcp and $3.07{}$~$\mu_B$
in fcc position.
The induced magnetic moments in the nearest-neighbor Ru atoms are much smaller than for Rh due to a smaller magnetic
susceptibility and amount to 0.03~$\mu_B{}$ and 0.07~$\mu_B{}$ for the hcp and fcc adsorption, respectively.

\begin{figure}
\includegraphics[width=0.9\linewidth]{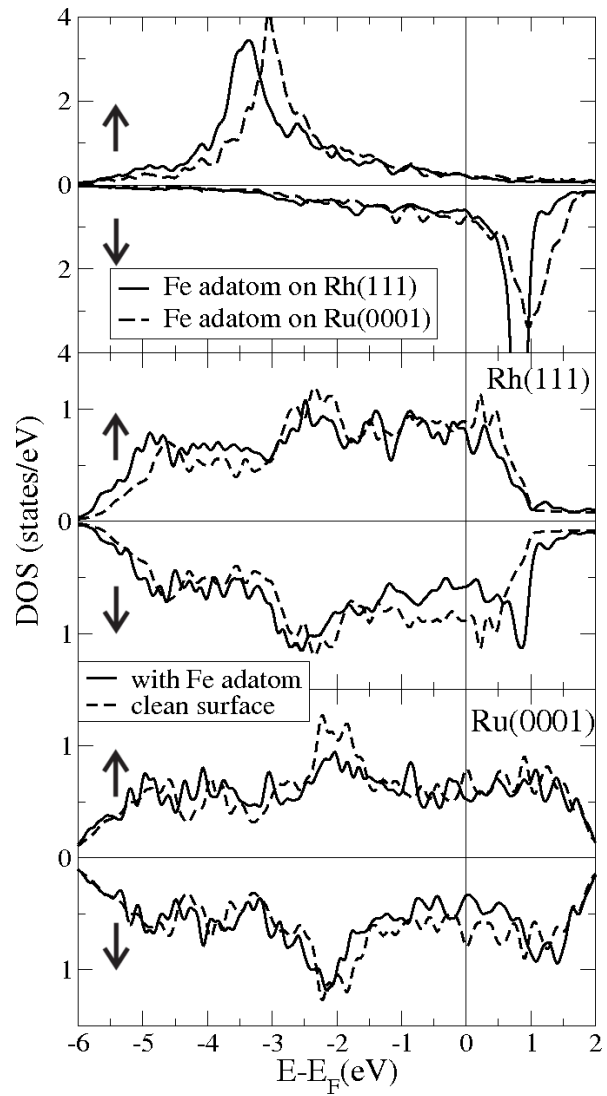}
\caption[fig1]{Local density of states (LDOS) for the Fe adatom on Rh(111) and Ru(0001) in the hcp adsorption site.
                The upper panel shows the LDOS
               of the Fe adatom on the two substrates. The middle and bottom panels display the LDOS of the pure Rh(111)
               and Ru(0001) surface, respectively and of the Rh and Ru surface atom adjacent to the Fe adatom.}
\label{fig:DOS_adatom}
\end{figure}

An analysis of the local density of states (LDOS) shown in Fig.~\ref{fig:DOS_adatom} further emphasizes the influence of the substrate.
For the Rh(111) surface the Fermi energy is shifted to higher energies within the $d$-band
compared to Ru(0001) due to the larger band filling.
In addition, the $4d$-band of Rh displays a smaller bandwidth than the $4d$-band of Ru which is caused by
the increased nuclear charge of Rh and the incomplete screening of the Coulomb potential due to the $d$-electrons.
These differences of the substrate density of states are also reflected in the hybridization with the Fe adatom. On
the Rh(111) surface the hybridization of the Fe $3d$-states with the substrate leads to pronounced peaks in the
majority and minority spin channel. The unoccupied minority $3d$-peak strongly interacts with the Rh states as seen in the
minority channel of the Rh LDOS at the same energy. In turn, there is a large induced magnetic moment of the Rh
surface atom. On the Ru(0001) surface the $3d$-peaks of the Fe adatom are broader due to a larger
overlap with the $4d$-states of Ru. However, a strong spin-polarization of the substrate does not occur. The
stronger Fe-Ru hybridization reduces the magnetic moment of the adatom as mentioned above.

\begin{figure}
\includegraphics[width=0.80\linewidth]{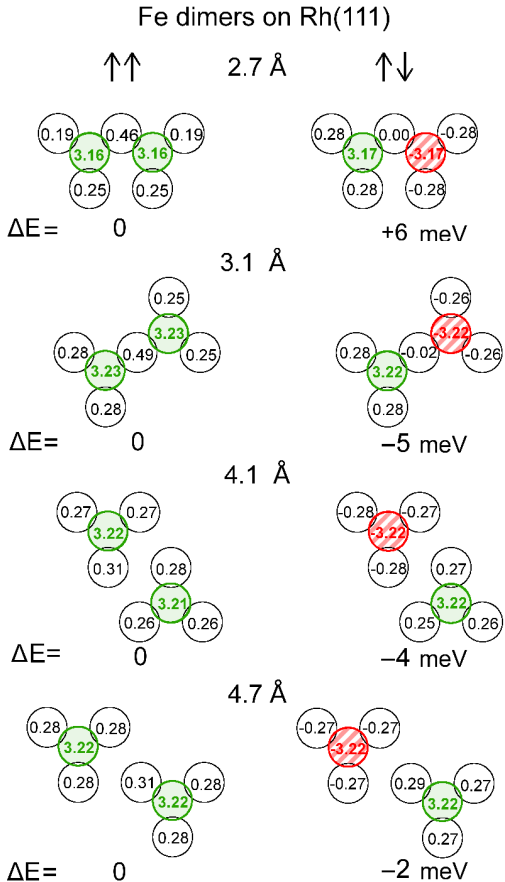}
\caption[fig1]{(color online) Geometric structure, magnetic moments, and energy differences for Fe dimers on Rh(111). 
               The spacing between the Fe atoms is given above each panel.
               Nearest-neighbor Rh surface atoms are shown by open circles
               while green and red circles denote Fe atoms of opposite magnetic moments.
               In all dimers one Fe atom sits in the energetically favorable
               hcp adsorption site. Left column shows the ferromagnetic ($\uparrow \uparrow$) state
               and the right column the antiferromagnetic ($\uparrow \downarrow$) configuration.
               The magnetic moment of the atoms is given in the circles in units of $\mu_{B\rm}{}$.
               The total energy in meV/Fe-atom is given at the bottom of each panel with respect to the ferromagnetic state.
               Note that these energy differences are not in all cases in one-to-one correspondence to the exchange constants given in
               Fig.~\ref{fig2} due to the interactions with atoms in adjacent unit cells.}
\label{fig1.1}
\end{figure}

\subsection{Dimers - magnetic moments and relaxations}
After discussing the structural, electronic, and magnetic properties
of the adatoms, we now turn to Fe dimers on the two substrates. We vary the spacing between the Fe
atoms in the dimers in our calculation from 2.7~{\AA} to 6.2~{\AA} within the $p(4\times4)$ unit cell
and allow fcc and hcp adsorption sites for the atoms.
As for the adatoms, the nearest-neighbor dimers prefer an hcp stacking, however, the energy gain with respect
to fcc sites is reduced to 65~meV/Fe-atom and 174~meV/Fe-atom on Rh(111) and on Ru(0001), respectively.
The data on the structural relaxations is summarized in table \ref{tab1}. Two magnetic configurations
were considered: a ferromagnetic ($\uparrow\uparrow$) and an antiferromagnetic ($\uparrow\downarrow$)
alignment of the Fe magnetic moments.

From table \ref{tab1} we find that on both surfaces the vertical relaxation is considerably reduced for
nearest-neighbor dimers with respect to the isolated Fe atoms.
This can be ascribed to the attractive interaction between the Fe atoms which display a lateral
relaxation towards each other.
For Fe dimers with an increasing spacing between the adatoms
the vertical relaxation eventually becomes equal to the relaxation of the Fe adatom.

On Rh(111) for dimers with a spacing of up to $d=3.1{}$ \AA\, the vertical relaxations are smaller for the
$\uparrow \uparrow$-state compared to the $\uparrow \downarrow$-state. This is related to the
Fe-Fe distance,  $d_{\rm{Fe-Fe}}{}$, which is about 0.1 \AA\, smaller in the $\uparrow \uparrow{}$ state
and reflects the larger attraction for ferromagnetically coupled Fe adatoms and a slightly weaker interaction
with the substrate.
A similar trend of the relaxations holds for dimers on Ru(0001).

\begin{figure}
\includegraphics[width=0.80\linewidth] {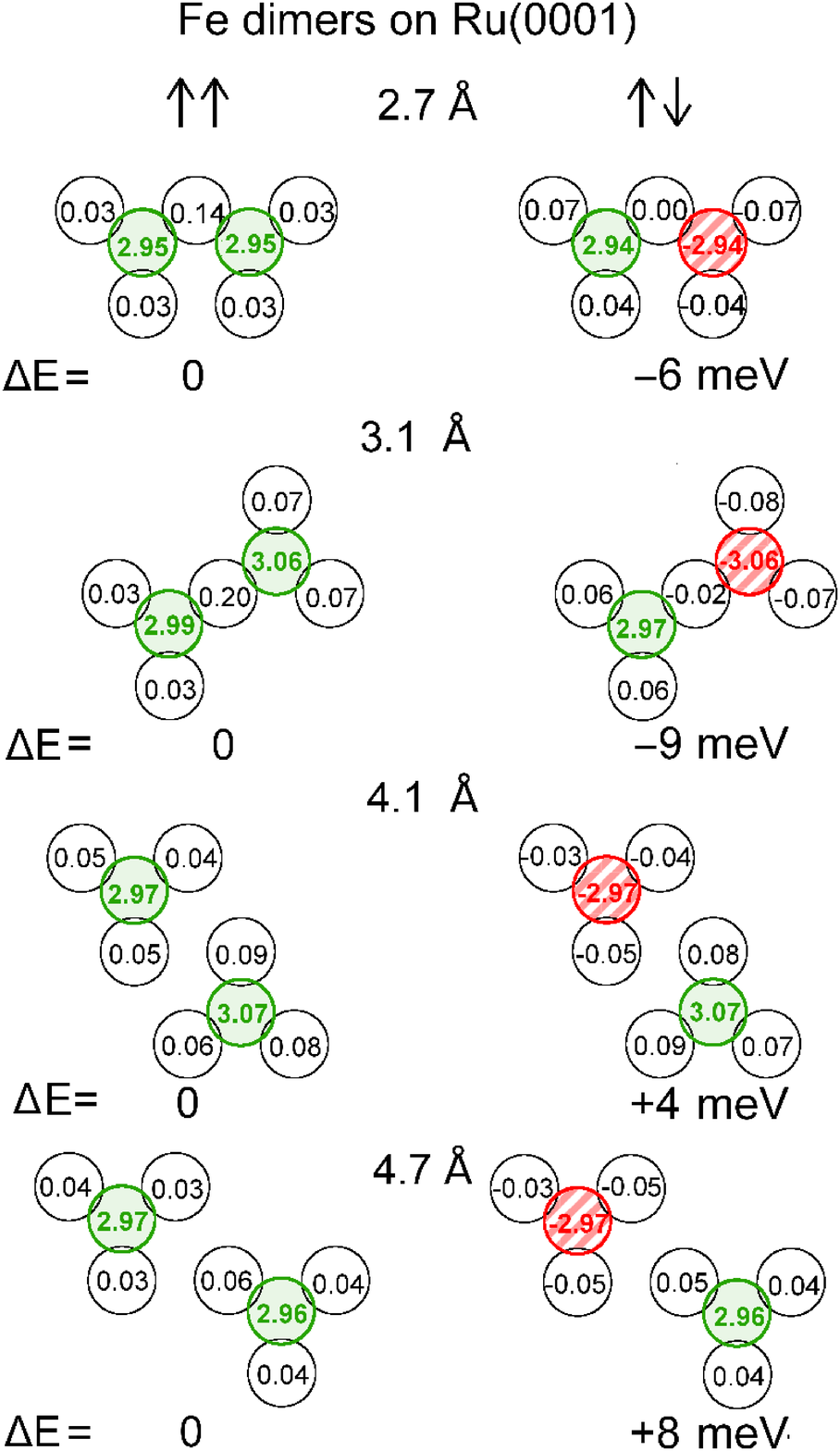}
\caption[fig1]{(color online) Geometric structure, magnetic moments, and energy differences for Fe dimers on Ru(0001).
               Refer to the caption of Fig.~\ref{fig1.1} for details.}
\label{fig1.2}
\end{figure}

Now we turn to the magnetic properties of the dimers presented in Fig.~\ref{fig1.1} for the Rh substrate
and in Fig.~\ref{fig1.2} for the Ru substrate.
On Rh(111) (Fig. \ref{fig1.1}) the magnetic moment of the Fe
atom is around $3.22{}$ $\mu_B$ in both hcp and fcc stacking except for the nearest-neighbor dimers,
where the moment is reduced by about $0.06{}$ $\mu_B$ due to the additional hybridization
between the Fe atoms.
The induced moment in the Rh surface atoms is on the order of 0.3~$\mu_B$
and rises for surface atoms with two Fe neighbors with parallel spin alignment
reaching values as high as about $0.5{}$ $\mu_B{}$.
On Ru(0001) (Fig. \ref{fig1.2}) the Fe magnetic moment depends on the adsorption site
and is about $2.97{}$ $\mu_B$ and $3.06$  $\mu_B$ for an hcp and fcc position, respectively.
This difference can be attributed to the large relaxation differences for atoms in hcp and
fcc adsorption sites on Ru(0001) as seen in table \ref{tab1}.
The induced magnetic moments of the Ru surface atoms are only on the order of 0.05 $\mu_B$
again with the exception of Ru surface atoms between two ferromagnetic Fe adatoms, where the induced magnetic moments reaches
$0.20{}$ $\mu_B{}$ for nearest-neighbor Fe dimers on fcc sites (not shown).

\subsection{Dimers - exchange interaction}
\begin{figure}

\includegraphics[scale=0.35]{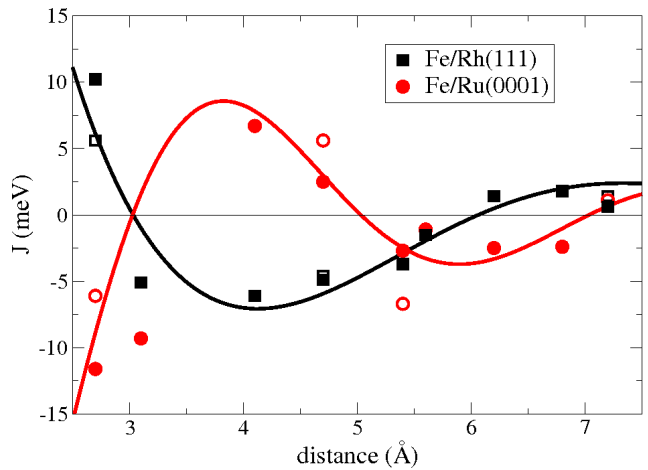}
\caption[fig1]{(color online) Exchange constants for Fe dimers on Rh(111) and on Ru(0001) as a function of the spacing between the
               atoms. Black squares represent results on the Rh surface and red circles on the Ru surface.
               Open symbols denote results for pure hcp adsorption sites and filled symbols mark pure fcc sites and mixed dimers.
               The fitting has been performed with an RKKY-like function
               (see text for details).
               }
\label{fig2}
\end{figure}

Next we focus on the exchange interaction in the Fe dimers on the two surfaces as a function of the interatomic separation.
In Fig. \ref{fig1.1} the energy differences are given between the FM and AFM configuration of the Fe dimers
up to fourth nearest neighbors on Rh(111). Note that these energy differences are not in all cases quantitatively equivalent
to the exchange constants shown in Fig.~\ref{fig2} as the interaction with Fe atoms in adjacent unit cells needs
to be taken into account. On Rh(111) the nearest-neighbor (NN) dimer is ferromagnetic (FM) and the ground state
switches to antiferromagnetic (AFM) for next-nearest neighbor dimers. Surprisingly, the absolute value of the two energy
differences and in turn the exchange interactions are very similar. This is in contrast to Fe dimers on other metal
substrates e.g. Cu(111) or Pd(111) on which the nearest-neighbor
exchange interaction clearly dominates~\cite{Mavro,FePd}. Due to the strong hybridization of the Fe adatoms with the Rh(111) surface the
nearest-neighbor exchange is strongly reduced as has been reported previously for Fe monolayer films on Rh(111)~\cite{Hardrat}. The
third and fourth NN exchange interaction remains antiferromagnetic before it switches back to FM at $d=6.2{}$ \AA \, (not shown).

Note that taking structural relaxations into account is of key importance for the determination of the exchange interaction in
this system. For
a nearest-neighbor Fe dimer on Rh(111) in the unrelaxed positions we obtain an energy difference of +57 meV/Fe-atom in favor of
the ferromagnetic state, while the value is +6 meV/Fe-atom for the relaxed dimer. Thus the exchange interaction is reduced by
one order of magnitude upon relaxation.

On Ru(0001), the exchange interaction between Fe atoms results in an AFM state for nearest and next-nearest neighbor
dimers and the exchange interactions are of similar
magnitude (Fig. \ref{fig1.2}). As on the Rh surface, the hybridization of the Fe adatoms with the substrate is strong,
which leads to the relatively small value of the NN exchange interaction. If one neglects the structural relaxation for
the nearest-neighbor dimer the energy difference even changes sign and amounts to $+30$~meV/Fe-atom instead of
$-6$~meV/Fe-atom, i.e.~the exchange
interaction becomes ferromagnetic. The exchange interaction for Fe dimers on the Ru surface
oscillates slightly faster with atom separation than on Rh(111) and changes to FM at 4.1 \AA \, and back to AFM at 5.4 \AA.

From the total energy differences between the ferro- and antiferromagnetic state we have obtained the exchange constants
as a function of Fe atom separation by mapping the energies onto the classical Heisenberg model.
For the determination of the exchange constants we have taken the interactions with atoms in adjacent cells into account.
In order to obtain a better accuracy for the exchange constants we have combined results from dimers calculated in the $p(4\times4)$ and
$p(3\times3)$ unit cells.
We have started from
dimer calculations which lead to the formation of periodic structures in the $p(4\times4)$ and $p(3\times3)$ unit cell.
This means that for dimers in the $p(4\times4)$ unit cell periodic structures are formed for separations beyond 5.4~{\AA},
i.e.~linear or zigzag chains with a separation between the atoms corresponding to that chosen for the dimer.
Since the separation of the atoms within the unit cell coincides with that for the nearest neighbor atoms in the
adjacent unit cells, these calculations should yield the most unperturbed exchange constants.
In this process we have
neglected exchange interactions for separations beyond 7.2~{\AA}. The set of exchange constants which we
obtained from these calculations were then used to eliminate the interaction with atoms in the adjacent
unit cells for dimer calculations which do not lead to periodic structures. By applying such an iterative
approach more accurate exchange constants could be obtained.

In Figure \ref{fig2} we summarize the distance dependence of the exchange constants derived in this way
from the dimer calculations. The values of the exchange constants have been fitted with a RKKY-like
function~\cite{NOTE_RKKY}.
Best fits have been achieved assuming a dependence with the
inverse square of the Fe-Fe separation, i.e.~$d^{-2}$, consistent with the surface geometry:
\[f(d)\propto \frac{\sin(2k_{\rm{F}}d+\phi)}{(2k_{\rm{F}}d)^2}.\]

The behavior of the RKKY-type oscillation is nearly inverted for Fe dimers on Ru(0001) compared
to Fe dimers on Rh(111) as seen in Fig.~\ref{fig2}.
For Fe dimers on Rh(111) the fit results in a Fermi wavelength of $k_{\rm F} \approx 0.51{}$ \AA$^{-1}$ and on Ru(0001) of
$k_{\rm{F}} \approx 0.78{}$ \AA$^{-1}$. We can identify a surface state near the Fermi level for Ru(0001) which might mediate
the interaction and is found in the bandstructure along the $\bar \Gamma \bar K$ high symmetry line.
The period of the oscillation on Rh(111) and Ru(0001) is comparably larger
than for example on Pt(111)\cite{Zhou} (0.3 \AA$^{-1}$) and on
Cu(111)\cite{Simon} (0.17 \AA$^{-1}$).

Reasons for the deviations of the fit from the calculated exchange constants may be the anisotropy of the
RKKY interaction due to the non-spherical Fermi surface and structural relaxations.
We attribute the larger deviations found for the exchange constants of dimers on the Ru surface
compared to Rh
to the large differences in vertical relaxations for Fe adatoms in hcp and fcc sites (cf.~table~\ref{tab1}).
Similar deviations from a perfect RKKY curve have been reported in a recent study~\cite{Zhou} that presented a direct comparison of
experimental and theoretical values of exchange constants for Co adatoms on Pt(111) obtained by spin-polarized STM and DFT calculations, respectively.

On both substrates there is a dependence of the exchange interaction
on the adsorption site of the Fe atoms, in particular for dimers in the nearest-neighbor configuration.
The FM and AFM coupling for nearest-neighbour dimers on Rh(111) and Ru(0001), respectively, is larger in fcc position as compared to hcp.
The discrepancy between dimers in hcp and fcc adsorption geometry are strongly reduced at larger separation on Rh(111).
However, this is not the case on Ru(0001), as clearly seen in Fig.~\ref{fig2} for larger distances, due to considerable differences in the structural relaxation for the two configurations (cf. table ~\ref{tab1}).

\section{Trimers and tetramers}
\label{sec:clusters}

\begin{table*}
\caption{\label{tab2}Average relative vertical relaxation $\bar \Delta_{12}{}$, relaxed Fe-Fe distances $d_{\rm{Fe-Fe}}{}$, and
                    energy differences with respect to the ferromagnetic solution of the three and four atom Fe clusters
                    and the full Fe monolayer on Rh(111) and Ru(0001).
                     For clarity only the average value
                     of $d_{\rm{Fe-Fe}}{}$ over all atoms in the tetramer is shown which is denoted as $\bar d_{\rm{Fe-Fe}}{}$.
                    The numbered circles depict the position of the Fe atoms. The energy differences are given for the fully
                     relaxed structures and in brackets for the unrelaxed clusters with respect to the ferromagnetic solution.
                     }
\begin{ruledtabular}
\begin{tabular}{lccccccccc}
  Fe on& &\multicolumn{4}{c}{Rh(111)} &\multicolumn{4}{c}{Ru(0001)} \\
  \hline
   &  & $ \bar \Delta_{12}{}$ (\%)   & \multicolumn{2}{c}{$ d_{\rm{Fe-Fe}}{}$ (\AA)}&$\Delta E{}$ (meV/Fe-atom)& $ \bar \Delta_{12}{}$ (\%) & \multicolumn{2}{c}{$ d_{\rm{Fe-Fe}}{}$ (\AA)} &$\Delta E{}$ (meV/Fe-atom)\\\hline

\multicolumn{2}{l}{Corner-shaped (hcp)}       &	    &        1-2   & 2-3   &  &   &  	1-2     &   2-3  &  \\\hline
& $\uparrow\uparrow\uparrow$ & $-17$  & 2.67& 2.63 &0 &$-14$   	&  2.67     &  2.65 &0        \\
\kreis{1} & $\downarrow\uparrow\uparrow$   & $-18$  & 2.76  & 2.60&$-5\ (+31)$ & $-14$	          & 2.71&	2.64 &$-6\ (+16)$\\
{ }{ }\kreis{2}{ }\kreis{3} & $\uparrow\uparrow\downarrow$  & $-18$ & 2.64  & 2.71 & $\ 0\ (+33)$  & $-14$		   &	2.67	 &2.70& $-7\ (+22)$ \\
 & $\uparrow\downarrow\uparrow$  & $-19$ & 2.75 & 2.70& $+7\ (+64)$ & $-14$       & 2.69  & 2.68 & $\ -15\ (+7)$  \\ \hline

\multicolumn{2}{l}{Linear (hcp)}       &	    &           1-2   & 2-3 &    &   &  	1-2     &   2-3&    \\\hline
& $\uparrow\uparrow\uparrow$& $-17$ & 2.67 & 2.67& 0 &$-14$ & 2.65 & 2.65 &0 \\
\kreis{1}{ }\kreis{2}{ }\kreis{3}&$\uparrow\uparrow\downarrow$  & $-18$  & 2.61 & 2.75& $-21\ (+12)$ & $-15$ & 2.64 & 2.72 & $-26\ (-1)$ \\
&$\uparrow\downarrow\uparrow$ & $-19$  & 2.71 & 2.71 &$-14\ (+39)$ &$-15$ & 2.69 & 2.69& $-28\ (+9)$\\\hline

\multicolumn{2}{l}{Triangle A (hcp)}       &	    &         1-2   & 1-3     &  &  &  	1-2     &   1-3 &   \\\hline
{ }{ }\kreis{3}& $\uparrow\uparrow\uparrow$  & $-16$   & 2.55 & 2.55&0 & $-12$ & 2.56 & 2.56 &0\\
\kreis{1}{ }\kreis{2}& $\uparrow\uparrow\downarrow$  & $-17$  & 2.56 & 2.66& $+55\ (+77) $&$-13$ & 2.63 & 2.66& $+32\ (+57)$ \\\hline
\multicolumn{2}{l}{Triangle B (hcp)}       &	    &         1-2   & 1-3 &    &    &  	1-2     &   1-3 &   \\\hline
\kreis{2}{ }\kreis{3}&$\uparrow\uparrow\uparrow$  & $-13$  & 2.50 & 2.50& 0 & $-10$ & 2.50 & 2.50 & 0 \\
{ }{ }\kreis{1}&$\uparrow\uparrow\downarrow$ & $-16$  & 2.52 & 2.65 & $+29\ (+92)$ & $-13$ & 2.56 & 2.64& $+2\ (+51)$ \\\hline

  Tetramer (hcp) && & \multicolumn{2}{c}{$ \bar d_{\rm{Fe-Fe}}{}$ (\AA)} & &  &  \multicolumn{2}{c}{$ \bar d_{\rm{Fe-Fe}}{}$ (\AA)}&	     \\\hline
\multirow{3}{*}{\minitab[l]{\kreis{3}{ }\kreis{4}\\{ }{ }\kreis{1}{ }\kreis{2}} } & $\uparrow\uparrow\uparrow\uparrow$      & $-15$  & \multicolumn{2}{c}{2.58}&0 & $-11$  & \multicolumn{2}{c}{2.56}&0 \\
 & $\uparrow\uparrow\downarrow\downarrow$    & $-16$ &  \multicolumn{2}{c}{2.60}&$+39\ (+78)$ &$-12$ &  \multicolumn{2}{c}{2.61}& $+23\ (+50)$ \\
 & $\uparrow\downarrow\downarrow\uparrow$    & $-16$  &  \multicolumn{2}{c}{2.59}&$+35\ (+92)$ & $-12$&  \multicolumn{2}{c}{2.59}& $+6\ (+44)$ \\\hline

\multicolumn{2}{l}{Monolayer (hcp)}       &	    &     \multicolumn{2}{c}{$ d_{\rm{Fe-Fe}}{}$ (\AA)} &    &   &   \multicolumn{2}{c}{$ d_{\rm{Fe-Fe}}{}$ (\AA)}&    \\\hline
& $\uparrow\uparrow$& $-6$ &  \multicolumn{2}{c}{2.70} & 0 &$-3$ & \multicolumn{2}{c}{2.70}  &0 \\
& $\uparrow\downarrow$& $-9$ &   \multicolumn{2}{c}{2.70}& $+19\ (+68)$ &$-7$ &  \multicolumn{2}{c}{2.70} &$-71\ (-9)$ \\\hline
  \hline

\multicolumn{2}{l}{Corner-shaped (fcc)}   &	    &         1-2   & 2-3  &   &  &  	1-2     &   2-3 &    \\\hline
&$\uparrow\uparrow\uparrow$		  & $-15$ 	& 2.59& 2.61 & 0 &$-10$&2.55&2.56&0 \\
\kreis{1} &$\downarrow\uparrow\uparrow$		  & $-16$ 	& 2.69& 2.61 &$+3\ (+30)$ & $-10$&2.63&2.56& $-4\ (+20)$\\
{ }{ }\kreis{2}{ }\kreis{3} &$\uparrow\uparrow\downarrow$		  & $-16$ 	& 2.58& 2.74 & $-3\ (+27)$& $-10$ &2.59&2.69&$-1$ \\
&$\uparrow\downarrow\uparrow$	  & $-17$ 	& 2.73 & 2.68 &$+12\ (+59)$& $-11$ &2.66&2.68&$-9\ (+21)$ \\\hline

\multicolumn{2}{l}{Linear (fcc)}  &	    &         1-2   & 2-3  &   &  &  	1-2     &   2-3 &   \\\hline
& $\uparrow\uparrow\uparrow$  & $-16$& 2.64     & 2.64& 0  & $-11$     & 2.60       & 2.60 &0    \\
\kreis{1}{ }\kreis{2}{ }\kreis{3}&$\uparrow\uparrow\downarrow$& $-17$ & 2.59     & 2.74& $-15\ (+7)$ &  $-14$    & 2.58  & 2.67 &$-12\ (+4)$     \\
&$\uparrow\downarrow\uparrow$& $-17$ & 2.69    & 2.69&$-5\ (+32)$  & $-12$     & 2.67       &  2.67&$-16\ (+11)$   \\\hline

\multicolumn{2}{l}{Triangle A (fcc)} 	      &	    &         1-2   & 1-3 &    &    &  	1-2     &   1-3 &   \\\hline
\kreis{2}{ }\kreis{3}& $\uparrow\uparrow\uparrow$  & $-11$  & 2.42   &  2.42 & 0  & $-8$  & 2.53     & 2.53 & 0     \\
{ }{ }\kreis{1}&$\uparrow\uparrow\downarrow$  & $-14$ & 2.48   &  2.63& +47 (+80) &$-10$    &  2.60    & 2.65 & $+31\ (+50)$ \\\hline
\multicolumn{2}{l}{Triangle B (fcc)} 	      &	    &         1-2   & 1-3    & &    &  	1-2     &   1-3  &  \\\hline
{ }{ }\kreis{3}&$\uparrow\uparrow\uparrow$  & $-14$  & 2.52   &  2.52 & 0  & $-4$   &    2.36    &   2.36 &0\\
\kreis{1}{ }\kreis{2} &$\uparrow\uparrow\downarrow$& $-16$  & 2.55  &  2.64&  $+56\ (+94)$ & $-9$   &    2.45    &   2.60 & $+39\ (+42)$\\\hline

  Tetramer (fcc) && & \multicolumn{2}{c}{$ \bar d_{\rm{Fe-Fe}}{}$ (\AA)} & &  &  \multicolumn{2}{c}{$ \bar d_{\rm{Fe-Fe}}{}$ (\AA)}	&     \\\hline

\multirow{3}{*}{\minitab[l]{\kreis{3}{ }\kreis{4}\\{ }{ }\kreis{1}{ }\kreis{2}} } & $\uparrow\uparrow\uparrow\uparrow$     & $-12$  &  \multicolumn{2}{c}{2.53}& 0& $-4$&   \multicolumn{2}{c}{2.46} &0\\

  &  $\uparrow\uparrow\downarrow\downarrow$    & $-14$  &  \multicolumn{2}{c}{2.57}& $+45\ (+82)$&$-9$ &  \multicolumn{2}{c}{2.58} &$+46\ (+52)$\\
  &  $\uparrow\downarrow\downarrow\uparrow$   & $-14$  &  \multicolumn{2}{c}{2.56}&$+45\ (+97)$ & $-9$ &  \multicolumn{2}{c}{2.57}&$+31\ (+47)$ \\
\hline

\multicolumn{2}{l}{Monolayer (fcc)}     &	    &     \multicolumn{2}{c}{$ d_{\rm{Fe-Fe}}{}$ (\AA)} &    &   &   \multicolumn{2}{c}{$ d_{\rm{Fe-Fe}}{}$ (\AA)}&   \\\hline
& $\uparrow\uparrow$& $-6$ &  \multicolumn{2}{c}{2.70} & 0 &$-0.4$ &  \multicolumn{2}{c}{2.70} &0 \\
& $\uparrow\downarrow$& $-8$ &  \multicolumn{2}{c}{2.70} & $+27\ (+79)$ &$-4$ &  \multicolumn{2}{c}{2.70}&$-8\ (+7)$ \\\hline

\end{tabular}
\end{ruledtabular}
\end{table*}

In this section we consider small Fe clusters of three and four atoms in different geometries adsorbed on
the Rh and Ru substrate. We compare the results with those for the adatom and dimer calculations in particular
with respect to the exchange interaction and the magnetic ground state.
First, an overview over the structural
relaxations is given and then the Fe magnetic moments and induced magnetic moments in the substrate are discussed.
Finally, the magnetic ground state of the different clusters and the exchange constants are analyzed in detail and
it is shown that it is crucial to take structural relaxations into account.

The investigated clusters can be divided into two groups: compact and open structures (for sketches see table \ref{tab2}).
The compact structures consist of tetramers and of two types of triangles which differ only by the adsorption geometry with
respect to the underlying surface. For triangle A there is a surface atom below the center of the triangle while
for B the central site is a hollow site of the substrate. The open structures are
the corner-shaped and the linear trimer.

\subsection{Structural relaxations}
For all types of clusters we consider hcp as well as fcc adsorption sites.
For both Fe clusters on Rh(111) and on Ru(0001) the hcp stacking is energetically preferred compared to fcc stacking.
On Rh(111), the energy difference between hcp and fcc stacking is 23~meV/Fe-atom for the compact trimer (triangle A)
which is already very close to the value of 13~meV/Fe-atom for the full monolayer. On Ru(0001), the corresponding energy
differences are 113 and 123~meV/Fe-atom. From a structural point of view,
the energetically most favorable state is triangle B, in which the Fe atoms relax stronger towards each other than
for triangle A. Triangle A and then the corner-shaped trimer
and the linear trimer exhibit an increasingly higher energy~\cite{NOTE1}.
As expected, we find that the more compact the structure,
i.e.~the more bonds with nearest-neighbor atoms can be formed, the more favorable it is.

\begin{figure}
\includegraphics[width=0.9\linewidth]{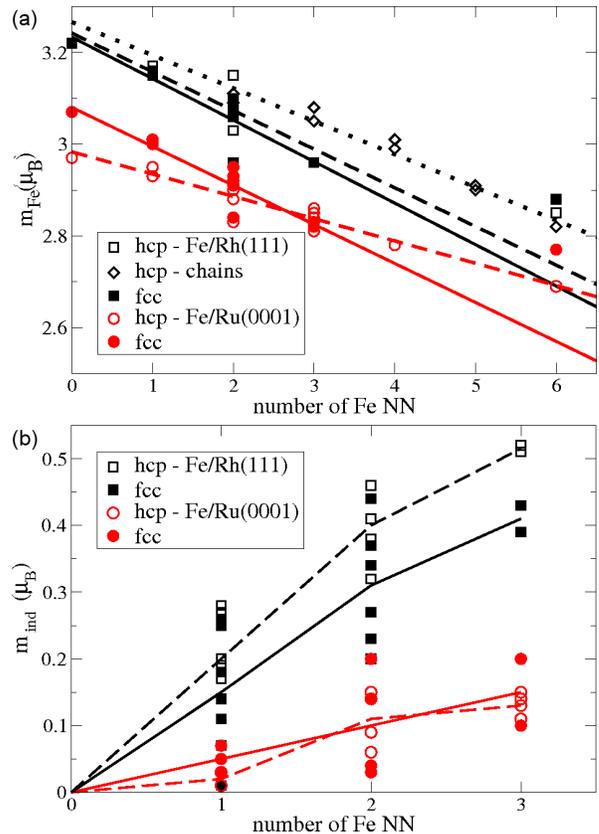}
\caption[fig5]{(color online) (a) Magnetic moments of Fe atoms in ferromagnetic clusters on Rh(111) and Ru(0001) as a function of the
number of nearest neighbor Fe atoms.
Solid, dashed, and dotted lines denote linear fits to the moment for fcc, hcp clusters, and hcp chains, respectively.
(b) Magnetic moments induced in the Rh(111) and Ru(0001) surface. Lines are guides to
the eye and obtained by choosing the average value of the induced moments at every
point. Values of the magnetic moments have been taken from nearest-neighbor dimers, compact and
open trimers, tetramers as well as the full monolayer. For the Ru surface also Fe pentamers
(section III.C) and for the Rh surface infinite three-strand chains (section \ref{sec:chains}) have been considered.}
\label{fig5}
\end{figure}

As we will see below structural relaxations are decisive for the magnetic ground state of many clusters.
In particular, if one assumes the Fe cluster atoms to be in perfect lattice positions of the substrate nearly all of the
considered clusters are ferromagnetic. However, the energetically most favorable magnetic state changes for many of them
upon structural relaxations.
Naturally, the interpretation of the energetics in terms of exchange constants is complicated due to the structural relaxations.

We discuss the cluster structures in terms of the average vertical relaxation of the adatoms and the Fe-Fe nearest
neighbor distances which can explain the observed trends. The deviations from these average values for specific
Fe atoms in a cluster do not modify the general picture.
The first point to notice is that clusters in fcc positions relax less along the direction perpendicular to the surface
than those in hcp positions in accordance with the results for the adatoms.
The average relative vertical relaxations, $\bar \Delta_{12}$, given in table \ref{tab2} show significant dependencies
on geometry and magnetic state of the clusters.
In general, there is a reduction of the vertical relaxation of cluster atoms with respect to the nearest-neighbor
dimers upon increasing the number of atoms in the cluster
due to hybridization and bonding with the additional Fe atoms.
As one would expect, the strongest effect occurs for the most compact structures, i.e.~for triangles A and B as well as
for the tetramer. In those cases
the relaxations are smaller by up to 7 \% on Rh(111) and up to 10 \% on Ru(0001) compared to NN dimers.
However, the relaxations are still much larger than for the full monolayer (cf.~table~\ref{tab2}).
The more open the geometry of a cluster is the stronger are the vertical relaxations. For the
linear and the corner-shaped trimer they are nearly the same as for the NN dimer (cf.~table~\ref{tab1}).

What cannot be seen from the average values of the relaxation is that edge atoms are subject to larger
vertical relaxation than central cluster atoms and that surface atoms are buckled.
These effects also depend on the magnetic coupling with neighboring Fe atoms and amount to a maximal change of 2\%.

We observe a similar trend on the cluster geometry for the values of the relaxed Fe-Fe distances, $d_{\rm{Fe-Fe}}{}$.
They are smallest for the compact structures, i.e.~triangle A and B and the tetramer, by up to 0.2 \AA \, on Rh(111) and
0.25 \AA \, on Ru(0001) with respect to the perfect spacing of 2.70~{\AA}.
Again the Fe-Fe distances for the open structures, i.e.~linear
and corner-shaped trimer, are closer to the NN dimer distances.
The influence of the magnetic state shows itself mainly by the fact that atoms with an opposite alignment of
magnetic moments hybridize less and thus the distance between them tends to be larger. These atoms are also
subject to larger vertical relaxation.
\begin{figure*}
\includegraphics[width=0.7\linewidth]{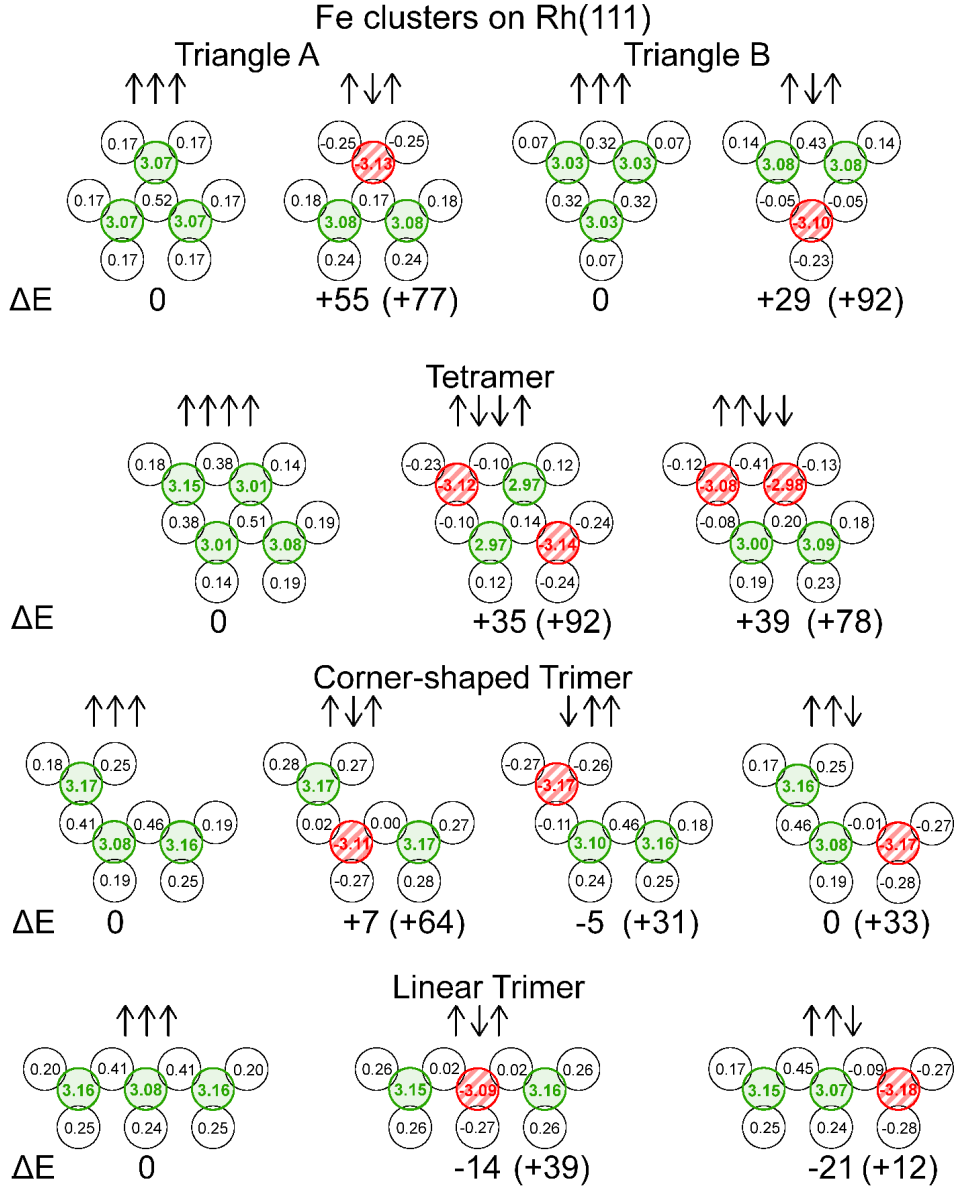}
\caption{(color online) Three and four-atomic Fe clusters on Rh(111) in hcp stacking for different magnetic configurations.
        Green and red circles denote atoms with magnetic moments in opposite directions. The arrows indicate the
        states as they are referred to in the text. The total energy differences are given in meV/Fe atom with
        respect to the FM solution (values for fcc stacking can be found in table \ref{tab2}).
        Values for structurally unrelaxed calculations with Fe atoms in the
        perfect Rh lattice positions are shown in parentheses. Calculations have been performed in the $p(4\times4){}$ unit cell.}
\label{fig:trimers_tetramers_Rh}
\end{figure*}
\subsection{Magnetic moments}
\begin{figure*}
\includegraphics[width=0.7\linewidth]{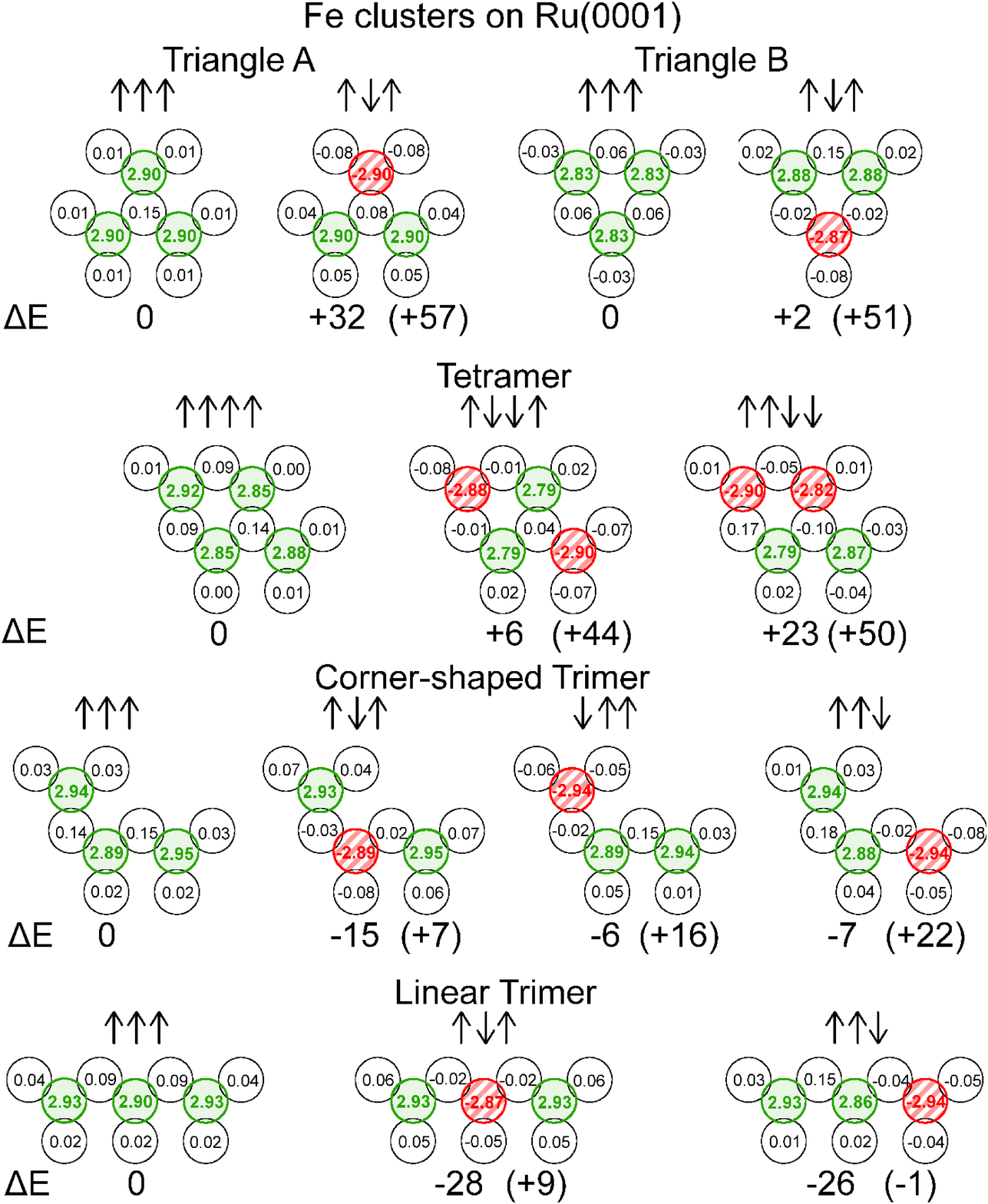}
\caption{(color online) Three and four-atomic Fe clusters on Ru(0001) in hcp stacking for different magnetic configurations.
        Refer to the caption of Fig.~\ref{fig:trimers_tetramers_Rh} for details.
}
\label{fig:trimers_tetramers_Ru}
\end{figure*}

We now turn to the magnetic properties of the clusters.
As seen in Fig.~\ref{fig5}(a),
we find a linear decrease of the size of the magnetic moments of the Fe atoms with the number of nearest neighbors.
The moments
have been obtained for clusters in the ferromagnetic state, however, the deviation from these values for antiferromagnetic
configurations is small (cf.~Figs.~\ref{fig:trimers_tetramers_Rh} and \ref{fig:trimers_tetramers_Ru}). The decrease
is due to hybridization with adjacent atoms which broadens the $3d$-states and leads to a smaller exchange splitting
and magnetic moment as expected from the Stoner model. The trend is in accordance with reports for Fe clusters
on other metal surfaces \cite{Mavro,Ir111}. On Rh(111) the magnetic moments of Fe cluster atoms in
hcp and fcc stacking are very similar. On Ru(0001), on the other hand, Fe atoms in fcc positions possess larger magnetic
moments than on hcp sites which can be explained based on the smaller vertical relaxations and in turn reduced hybridization
with the Ru surface atoms. This effect is most pronounced for Fe atoms in open structures having few nearest neighbors.
The variation of vertical relaxations with cluster size and shape are also the origin of the deviations from a linear
fit for Fe atoms with six nearest neighbors, i.e.~in the full monolayer. For Rh(111) we obtain similar vertical relaxations
for Fe atoms in infinite monoatomic, biatomic, and three-strand chains (discussed in section \ref{sec:chains})
and in the full monolayer and accordingly the magnetic moments reveal
a linear decrease which nicely matches the full monolayer.
Clusters in hcp stacking on the Ru(0001) surface also display a linear trend in agreement with the full monolayer.

The high magnetic susceptibility of Rh is immediately obvious from the large induced magnetic moments in the
Rh atoms that are nearest neighbors of Fe cluster atoms as shown in Fig.~\ref{fig5}(b). In general, the Rh magnetic moments
increase for atoms with more nearest-neighbor Fe atoms of the same spin alignment.
However, there is a large spread in the values depending on the exact cluster geometry.
Rh surface atoms which have neighboring
Fe atoms with opposite magnetic moments are naturally polarized much less due to a partial compensation (cf.~Fig.~\ref{fig:trimers_tetramers_Rh}).
For Rh atoms with three Fe neighbors, the magnetic moments are larger for Fe adatoms in hcp sites than in fcc sites.
This reflects the slightly stronger relaxations for Fe atoms in hcp sites. The maximum values of the induced Rh
magnetic moments vary between 0.4 and 0.5 $\mu_B{}$.

Ru possesses a much smaller susceptibility than Rh but a similar dependence of induced spin-polarization on the
number of nearest-neighbor
Fe atoms is observed in Fig.~\ref{fig5}(b). There one can also see that the maximum values of the induced magnetic moments
are on the order of 0.2 $\mu_B{}$.
In contrast to the Rh(111) surface, the spin-polarization is stronger for Ru atoms with neighboring Fe adatoms in fcc sites
than in hcp sites.
This can be understood based on the larger Fe moments in fcc adsorption sites which results from their
smaller inward relaxation, compared to the very similar values for Fe atoms in hcp and fcc sites on Rh(111).

\subsection{Magnetic ground states}
Now we discuss the magnetic ground states of the trimers and tetramers. Figure \ref{fig:trimers_tetramers_Rh}
shows the geometry and the different collinear magnetic states which we considered for the Fe clusters on Rh(111).
In the figure the energy differences are given for the clusters in hcp stacking while those for fcc stacking
can be found in table \ref{tab2}. One can immediately see that the geometry is decisive for the magnetic ground state.
The compact clusters, i.e.~triangles A and B as well as the tetramer, possess a ferromagnetic ground state
with a significant energy difference compared to antiferromagnetic states. In contrast, for the open structures,
i.e.~the corner-shaped and the linear trimer, the antiferromagnetic $\downarrow \uparrow \uparrow$-state is energetically
most favorable. This finding is very surprising at first glance because the nearest-neighbor exchange
is expected to be ferromagnetic based on our dimer calculations (cf. Fig. \ref{fig2}). For the corner-shaped
trimer, the antiferromagnetic configuration can still be understood based on ferromagnetic nearest-neighbor
exchange. However, there are two inequivalent NN exchange couplings and one of them is reduced significantly
as will be discussed in more detail at the end of this section.
Therefore, the second nearest-neighbor exchange comes into play and leads to the $\downarrow \uparrow \uparrow$-state.
For the linear trimers, on the other hand, the nearest-neighbor exchange interaction changes sign
due to weakened direct Fe-Fe exchange as we will show at the end of this section.

As in the calculations for the Fe dimers, we find
that structural relaxations have a significant influence on the energy differences between magnetic configurations.
For the compact clusters on Rh(111), we observe that upon relaxation the energy differences between ferro- and antiferromagnetic
states are greatly reduced (cf.~Fig.~\ref{fig:trimers_tetramers_Rh}).
The largest changes occur for triangle B in hcp stacking where the energy
difference drops from +92~meV/Fe-atom to +29~meV/Fe-atom and for the tetramer in hcp stacking with a change from
+92~meV/Fe-atom to +35~meV/Fe-atom. The origin of such a strong reduction lies in the hybridization
with the substrate which leads to a weakened nearest-neighbor FM exchange in agreement with our observations for the
nearest-neighbor dimers. However, the vertical relaxations of the triangles and tetramers is by about 2 to 5 \%
smaller compared to the NN dimer and the spacing between the Fe atoms is reduced by about 0.1 {\AA}. Therefore,
the direct ferromagnetic exchange between Fe atoms is strengthened and the energy gain of the ferromagnetic state is
by one order of magnitude larger than for the dimers (cf.~Fig.~\ref{fig1.1}).

For the open structures on Rh(111), the effect is even more dramatic.
For all linear and corner-shaped trimers the magnetic ground state changes from ferromagnetic without relaxations to
an antiferromagnetic one after full structural relaxation. The energy gain due to the interaction in favor of
antiferromagnetic states is of similar magnitude as for the compact structures (cf.~Fig.~\ref{fig:trimers_tetramers_Rh}).
For the corner-shaped trimers
this leads to an energetically slightly more favorable $\downarrow \uparrow \uparrow$-state by 3 and 5 meV/Fe-atom
in fcc and hcp stacking, respectively, while for the linear trimers these energy differences are 15 and 21 meV/Fe-atom.
The difference between hcp and fcc stacking can be explained by the two non-equivalent NN exchange couplings, which are
interchanged between hcp and fcc stacking (cf.~Fig.~\ref{fig2}).
To check the influence of the unit cell size, the corner-shaped trimer in hcp stacking has been calculated in the $p(3\times3){}$
and $p(5\times5){}$ unit cell for comparison.
The energy differences between the FM and the $\downarrow\uparrow\uparrow{}$
state amount to $-10{}$ meV/Fe atom and $-7{}$ meV/Fe atom, respectively, compared to $-5{}$ meV/Fe atom in the $p(4\times4){}$
unit cell. This shows that the size of the unit cell do not influence the magnetic ground state.

A similar evolution of magnetic states is observed in Fig.~\ref{fig:trimers_tetramers_Ru} for Fe clusters on Ru(0001).
However, for the compact clusters the energy
difference between ferro- and antiferromagnetic states is smaller since the NN exchange interaction has an even stronger
antiferromagnetic tendency as observed already for the dimers.
In particular, for hcp stacking the FM and AFM states
of triangle B and the tetramer are nearly degenerate with differences of 2 and 6 meV/Fe-atom, respectively.
Nevertheless, all compact clusters are still ferromagnetic in contrast to our expectation based on the dimer results.
This change to ferromagnetic exchange interaction in compact clusters is due to the increased number of Fe
nearest neighbors and different vertical and lateral structural relaxations which
modifies the competition between the direct Fe-Fe exchange and the hybridization with the Ru substrate atoms.
The vertical relaxations are considerably reduced for compact clusters on Ru(0001) by about 3 to 7 \% compared to the
values of the NN dimers and the Fe-Fe distances are lower by about 0.1 {\AA} (cf.~tables~\ref{tab1} and \ref{tab2}).
Therefore, the exchange interaction
in the compact clusters is still ferromagnetic in contrast to what we would have expected by naively using the NN
antiferromagnetic exchange interaction from the dimer calculations.

\begin{figure}
\includegraphics[width=0.75\linewidth]{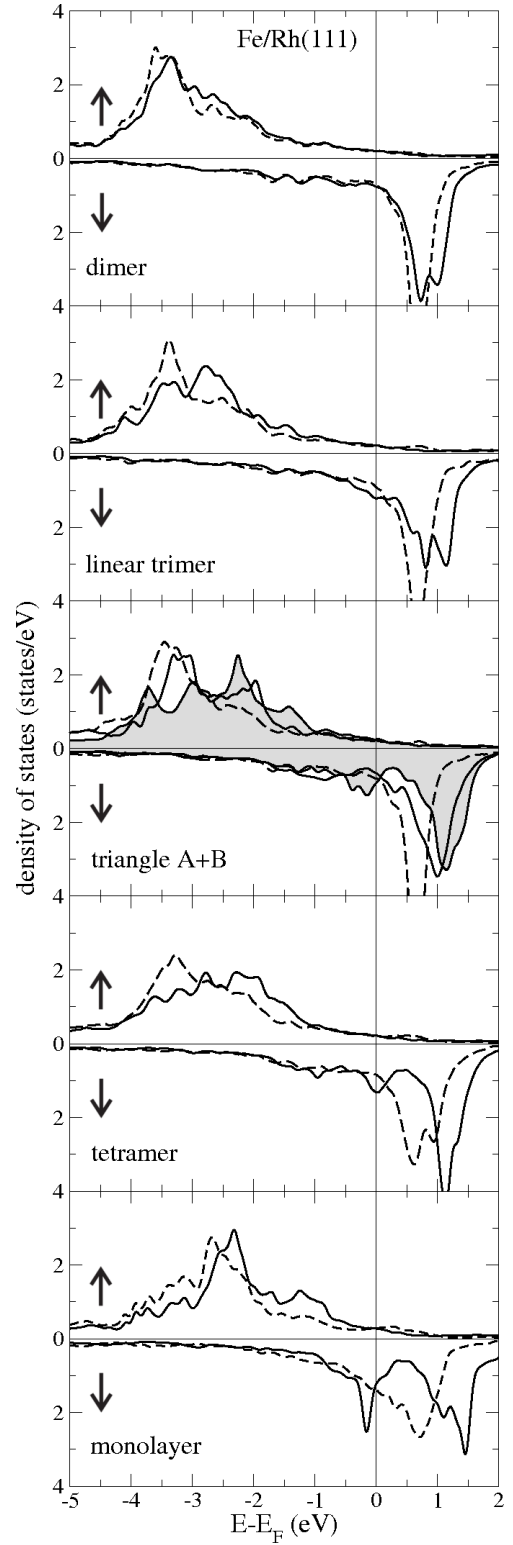}
\caption[fig_FeRh_DOS]{Local density of states of Fe clusters on Rh(111) in hcp stacking.
From top to bottom the panels show the LDOS of a dimer, linear trimer, compact trimer (triangle A/B),
a tetramer, and the full monolayer. In every panel spin-up and -down channels are shown and solid
and dashed lines denote the ferro- and antiferromagnetic state, respectively. For the linear trimer
and triangle A the $\uparrow\downarrow\downarrow$-state is shown and the LDOS of the Fe atom with the
$\uparrow$ moment is given.
For triangle B only the LDOS of the FM state is shown by a gray filled area.
}
\label{fig:FeRh_DOS}
\end{figure}

In order to further illustrate the influence of lateral and vertical relaxations on the exchange interaction in the compact Fe clusters,
we have artificially constrained the relaxations in a calculation for triangle B in hcp stacking on Ru(0001). If we allow only
a relaxation in the vertical direction keeping the Fe-Fe distances from the ideal lattice positions, the energy difference changes
from +2~meV/Fe-atom indicating a ferromagnetic state to $-9$~meV/Fe-atom in favor of the antiferromagnetic state. We conclude
that the hybridization with the substrate drives the cluster towards an antiferromagnetic state while the direct ferromagnetic
exchange interaction between the Fe atoms in the cluster is strengthened by lateral relaxations.

For the corner-shaped and the linear trimer on Ru(0001), antiferromagnetic states are also more favorable, however,
consistent with the AFM nearest-neighbor exchange observed for the dimers, the $\uparrow\downarrow\uparrow$-state
is preferred. As for the compact structures, the energy gain in favor of antiferromagnetic states is larger than
on the Rh substrate so that on Ru(0001), also the corner-shaped trimers possess a clear energy gain of the
antiferromagnetic configuration of 15 meV/Fe-atom.
For the linear trimer in hcp stacking on Rh(111) calculations have been performed for the energy difference between the
FM and the $\uparrow\uparrow\downarrow{}$ state in the $p(5\times5){}$ unit cell, which leads to a value of $-22{}$ meV/Fe atom
close to the $-21$~meV/Fe-atom obtained in the $p(4\times4){}$ unit cell (cf.~Fig.~\ref{fig:trimers_tetramers_Ru}).
This also confirms that the chosen unit cell
is sufficiently large so that interactions with atoms in the adjacent unit cell are small and do not alter our conclusion
on the magnetic ground state configuration.

\subsection{Local density of states}
It is instructive to compare the local density of states for the different clusters. In Fig.~\ref{fig:FeRh_DOS} the
LDOS is shown for Fe clusters on Rh(111) ranging from the dimer to the full monolayer. For the dimer, we observe a
very sharp peak of the $3d$-states in particular for the minority spin channel similar to that of the adatom
(cf.~Fig.~\ref{fig:DOS_adatom}).
In the ferromagnetic configuration, this state can split due to bonding between the two Fe atoms while it remains
sharp in the antiferromagnetic state. Upon increasing the number of nearest-neighbor Fe atoms in the cluster, the
density of states splits and broadens further in the ferromagnetic state due to additional hybridization. In the
antiferromagnetic configuration, this effect is much weaker since the $3d$-peaks of adjacent Fe atoms are in
opposite spin channels and only the tails overlap. As the number of nearest neighbors in the cluster starts to
increase for the triangle and the tetramer to two and three, respectively, the LDOS splits into more peaks and
the unoccupied antibonding minority states shift to higher energies. On the other hand, bonding states move below the Fermi
energy leading to a smaller magnetic moment. For the tetramer the LDOS already starts to resemble the full monolayer
in which every Fe atom has six nearest-neighbors and the splitting of the majority and minority
spin LDOS into bonding and antibonding states is fully developed.

\begin{figure}
\includegraphics[width=0.75\linewidth]{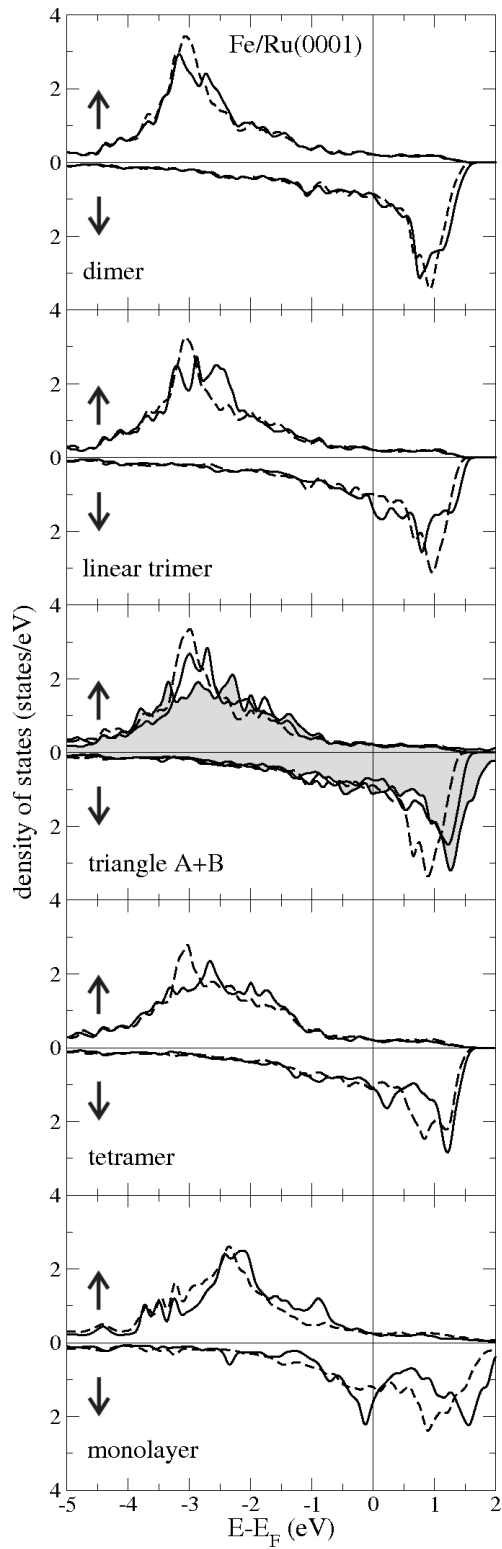}
\caption[fig_FeRu_DOS]{Local density of states of Fe clusters on Ru(0001) in hcp stacking.
From top to bottom the panels show the LDOS of a dimer, linear trimer, compact trimer (triangle A/B),
a tetramer, and the full monolayer. In every panel spin-up and -down channels are shown and solid
and dashed lines denote the ferro- and antiferromagnetic state, respectively. For the linear trimer
and triangle A the $\uparrow\downarrow\downarrow$-state is shown and the LDOS of the Fe atom with the $\uparrow$ moment
is given. For triangle B only the LDOS of the FM state is shown by a gray filled area.
}
\label{fig:FeRu_DOS}
\end{figure}

For the Fe clusters on Ru(0001), we find a similar trend of the LDOS as seen in Fig.~\ref{fig:FeRu_DOS}. In
comparison to the Rh substrate, the unoccupied minority peak in the $d$-states is pushed to higher energies
and accordingly the tail of the peak also moves. For the linear trimer and the tetramer this leads to a shift
of a small peak which is at the Fermi energy for Fe on the Rh surface to being slightly above the Fermi energy on
the Ru substrate. Since the exchange interaction for Fe depends sensitively on the level of the Fermi energy with
respect to the minority $d$ band \cite{Mavro,Lukashev2013} such changes can be responsible for the tendency to Fe
clusters on Ru to couple antiferromagnetically.

The crucial influence of the local geometry on the exchange interaction is evident from a comparison of triangles
A and B. The only difference between these two cluster geometries is that triangle B does not possess a
surface atom below its center and
less NN surface atoms. On both, Rh(111) and Ru(0001), this leads to a drastic reduction of the energy difference
between the ferro- and antiferromagnetic state by about 30 meV/Fe-atom for clusters in hcp stacking. For triangles
in fcc stacking, there is still a difference, however, it is only about 9 meV/Fe-atom (cf.~table \ref{tab2}).
The modified hybridization between the cluster atoms and the substrate is also observed in the local DOS for
these two types of triangles as shown in Figs.~\ref{fig:FeRh_DOS} and \ref{fig:FeRu_DOS}.
For triangle A in hcp stacking on Rh(111) one see one main peak at $-3.25{}$~eV and a plateau next to it in the majority
DOS. However, this structure splits into four peaks in the cluster geometry of triangle B which indicates the
major influence of the different stacking on the hybridization with the substrate.
In the minority states, we also observe changes between the two triangular configurations, in particular,
just below the Fermi energy a peak structure appears in the cluster with the triangle B geometry. As the
electronic states and hybridization at the Fermi energy are crucial for this system such modifications
can result in the energy differences between the magnetic states.
On Ru(0001) we find quite similar structures in the local DOS for the two cluster configurations which is
in accordance with the similar changes of the exchange interaction.
\subsection{Exchange interactions}
\begin{table}
\caption{\label{tab3}Comparison of exchange constants derived from Fe dimers, trimers, and tetramers
            on Rh(111) and on Ru(0001) in hcp and fcc stacking. The exchange constants $J_1$, $J_2$, and
            $J_3$ are defined as first, second, and third nearest-neighbors considering either only hcp
            or only fcc sites. Positive and negative signs denote ferro- and antiferromagnetic exchange
            coupling, respectively. Note that for the corner-shaped trimer there are two inequivalent
            nearest-neighbors and thus also two values for $J_1$.}
\begin{ruledtabular}
\begin{tabular}{lcccccc}
  \hline
   Fe on & \multicolumn{3}{c}{Rh(111)} & \multicolumn{3}{c}{Ru(0001)} \\
   & $J_{1}$ & $J_{2}$ &$J_{3}$  & $J_{1}$  & $J_{2}$ & $J_{3}$ \\

hcp & \multicolumn{3}{c}{} & \multicolumn{3}{c}{}  \\\hline
dimer & 5.6  & $-4.6$& $-3.7$ &$-6.1$ & 5.6  & $-6.7$ \\
triangle A  &   41.3& -  & -  & 24.0    & -  & - \\
triangle B  &   21.8  & -  & - &   2.0    & -  & - \\
tetramer    &   18.0  & 25.0 & - & 3.0 & 37.0 & - \\
corner-shaped & 8.8/1.0 & $-8.8$& -& $-12.1$/$-10.7$  & 1.9   & - \\
linear & $-10.3$& - & $-22$& $-20.9$  & - & $-17.9$ \\
            &        &      &   &        &    &   \\
fcc & \multicolumn{3}{c}{} & \multicolumn{3}{c}{}  \\ \hline
dimer & 10.2& $-4.9$ & $-3.7$ & $-11.6$ & 2.5 & $-2.7$ \\
triangle A  &  35.3 & -  & -  & 23.3    & -  & - \\
triangle B  &  42.0  & -  & -  & 29.3   & -  & - \\
tetramer    &   23.0  & 22.0 & - & 16.0 & 45.0 & - \\
corner-shaped & 13.3/5.3& $-9.1$ & - & -  & - & - \\
linear & $-3.6$ & - & $-19.7$ & $-11.7$ & - & $-7.6$ \\
\end{tabular}
\end{ruledtabular}
\end{table}

Finally, we analyze the total energies for the different magnetic configurations of the clusters on the two substrates
in terms of exchange constants by mapping them to a Heisenberg model. The results are summarized in table \ref{tab3}.
We find large changes in exchange constants for different geometries consistent with previous studies of other
systems reported in Refs.~\onlinecite{Mavro,Ir111}.

Taking a look at the compact structures first, we observe that for triangles A and B in hcp stacking on the Rh(111)
surface the nearest-neighbor exchange constants amount to about $+41$~meV and $+22$~meV, respectively. In the fcc
configuration, the exchange is on the same order of magnitude as seen in table \ref{tab3}. These values are
about four to eight times larger than those obtained for the nearest-neighbor dimers. This finding underlines
our statement above that the ferromagnetic exchange is much strengthened in the compact clusters due to smaller
vertical relaxations, smaller Fe-Fe spacing, and more nearest Fe neighbors which increases the ferromagnetic direct
exchange between the Fe atoms. On the Ru(0001) surface, the effect is very similar, but the values of the exchange constants
are small. The sign of the exchange interaction, however, has changed from antiferromagnetic for nearest-neighbor dimers to
ferromagnetic exchange in the triangles. Thus the direct ferromagnetic exchange between the Fe atoms in the compact
clusters prevails over the tendency towards antiferromagnetic coupling induced by the hybridization with
the Ru substrate. Interestingly, we find that triangle B in hcp stacking exhibits only a very small NN ferromagnetic
exchange. This is in accordance with the much reduced value of the exchange interaction for the same cluster
geometry on the Rh(111) surface. Thus the details of the cluster geometry and hybridization with the
substrate turn out to be decisive for the coupling in these systems.

For tetramers, we obtain a similar magnitude of the exchange coupling, however, the nearest-neighbor exchange is
reduced with respect to the triangular clusters. The coupling with the second nearest neighbor within the
tetramer is very strong and even exceeds the NN exchange interaction in most cases. For hcp stacking,
$J_1$ is actually quite close to the value found for triangle B while for fcc stacking it is closer
to triangle A.

For the open cluster geometries, i.e.~the corner-shaped and the linear trimer, the situation is quite different
than for the compact structures. For the corner-shaped trimers, we obtain two inequivalent nearest-neighbor
exchange constants as well as the next-nearest neighbor exchange.
From table \ref{tab3}, we find that for the corner-shaped trimers on both substrates and in
both stackings the sign of the nearest and next-nearest neighbor exchange interaction is the same as for the dimers.
Even the magnitude is comparable which indicates that the structural relaxations (cf.~table \ref{tab2})
and the electronic structure,
i.e.~the hybridization between the atoms in the trimer as well as with the substrate, are relatively similar.

For the linear trimers the situation is more complex as observed in table \ref{tab3}. We can extract a value of the
nearest-neighbor exchange, $J_1$ and the third nearest-neighbor exchange interaction, $J_3$. On the Rh(111) surface,
both exchange couplings in the linear trimer are antiferromagnetic and the exchange with the second neighbor is even
larger. In comparison with the dimers, there is a change of sign of $J_1$ which was ferromagnetic in the dimer and there
is a huge increase of $J_3$ which is also antiferromagnetic for dimers. Apparently, the antiferromagnetic exchange
is much favored as we build a linear atomic chain of Fe atoms on the Rh surface. This conclusion is further strengthened
from our results on infinite chains as discussed in section III.D. The electronic structure of the linear trimers and
infinite chains is also quite similar as observed in the local DOS and deviations significantly from that of the dimers
(cf.~Fig.~\ref{fig:FeRh_DOS}).
Linear Fe trimers on Ru(0001) display a similar trend towards an increase of antiferromagnetic exchange constants
between nearest and next-nearest neighbors in the trimer. As on Rh(111) the effect is even more pronounced in hcp stacking
and $J_3$ is of comparable size as $J_1$.

By comparing the Heisenberg exchange constants derived from the dimer and from the three atomic cluster calculations in table \ref{tab3},
it shows that on Rh(111) $J_{1}{}$ becomes smaller when moving from the dimer to the corner-shaped trimer and eventually AFM for the
linear trimer. On Ru(0001) $J_{1}{}$ is already AFM for the nearest neighbor dimer and becomes more strongly AFM when moving to the
open structured trimers. This is consistent with the observations in Ref.~\citenum{Mavro} that nearest-neighbor ferromagnetic exchange
tends to be weakened for linear trimers compared to the corner-shaped ones. This effect can be understood based on the symmetry of the
$3d$-orbitals of the adatoms and the splitting of the density of states upon hybridization~\cite{Mavro}.

\subsection{Open tetramer structures}
\label{sec:open_tetramers}
\begin{figure*}
\includegraphics[width=0.7\linewidth]{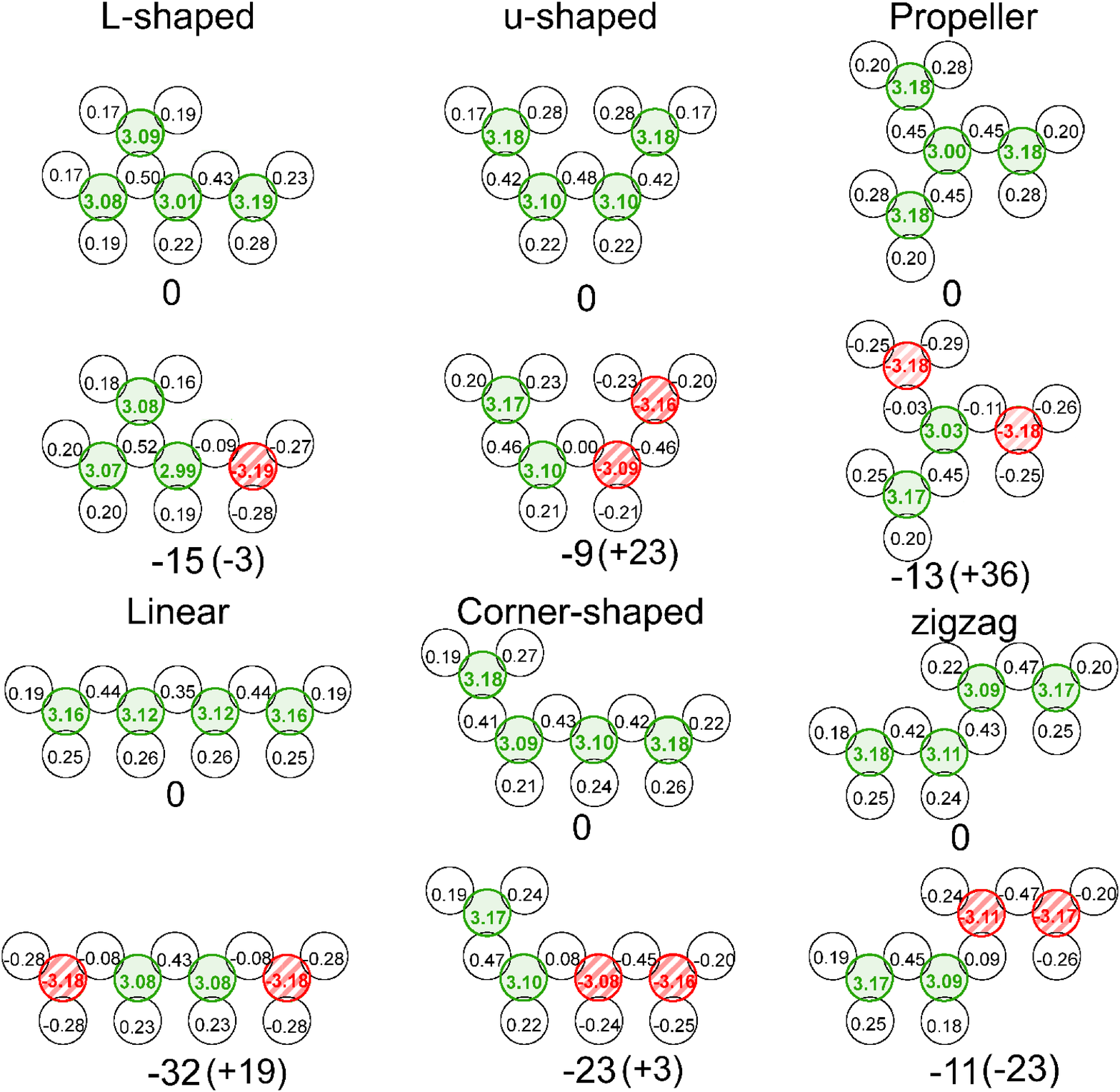}
\caption{(color online) Four-atom Fe clusters on Rh(111) in hcp stacking for different magnetic configurations.
        Green and red circles denote atoms with magnetic moments in opposite directions. The total energy differences are given in meV/Fe atom with
        respect to the FM solution. Values for structurally unrelaxed calculations with Fe atoms in the
        perfect Rh lattice positions are shown in parentheses. Calculations have been performed in the $p(5\times5){}$ unit cell.
        Besides the depicted states, the following states with the respective energy difference to the FM state given in parentheses
        have been calculated for clusters on Rh(111). The arrows indicate the orientation of the magnetic moment of the:
L-shaped: $\uparrow\uparrow\downarrow\downarrow{}$ (+19 meV/Fe-atom); u-shaped: $\uparrow\downarrow\uparrow\downarrow{}$ (+1 meV/Fe-atom);
linear: $\uparrow\downarrow\uparrow\downarrow{}$ ($-31$ meV/Fe-atom), $\uparrow\uparrow\downarrow\downarrow{}$ ($-31$ meV/Fe-atom);
zigzag: $\uparrow\downarrow\uparrow\downarrow{}$ ($-1$ meV/Fe-atom).}
\label{fig:tetramers_Rh}
\end{figure*}

To complete the picture of the tetramers we briefly summarize the results for open tetramers and structures which are geometrically
intermediate in between the compact and the open structures.
The favorable magnetic states which we found for these tetramers are consistent with the picture which we have developed
in section III.C for the open trimers, i.e.~all of them possess a compensated antiferromagnetic ground state if we take
structural relaxations into account.

\begin{figure*}
\includegraphics[width=0.7\linewidth]{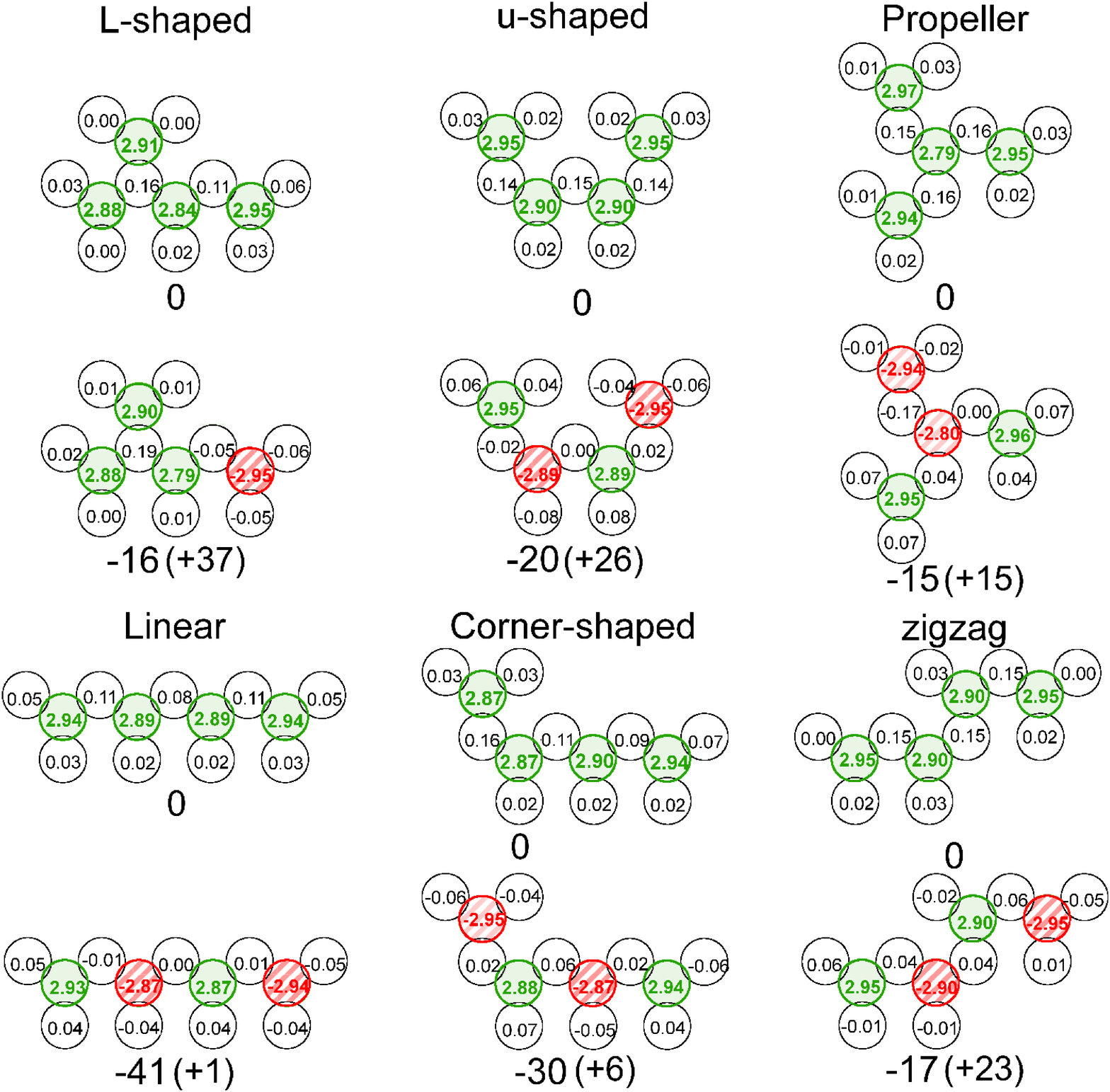}
\caption{(color online) Four-atom Fe clusters on Ru(0001) in hcp stacking for different magnetic configurations.
        Refer to the caption of Fig.~\ref{fig:tetramers_Rh} for details. }
\label{fig:tetramers_Ru}
\end{figure*}

In Figs.~\ref{fig:tetramers_Rh} and \ref{fig:tetramers_Ru} the
comparison of the ferromagnetic and compensated magnetic states is presented in terms of the total energy differences.
It is computationally very demanding to determine the collinear magnetic ground state for the tetramers since there are
many inequivalent magnetic states which would all need to be individually relaxed concerning their structure.
Therefore we have only calculated the magnetic states that are most likely to be the ground state based on the behavior observed for the
trimers. For some geometries of the clusters on Rh(111) we also tested further magnetic configurations which are not shown here but mentioned in the figure caption.
Even if the considered compensated states are not the magnetic ground state, one can still conclude the tendency of different
geometries towards the ferromagnetic or a compensated state based on our results.

For Fe tetramers on Rh(111) we have considered the $\uparrow\uparrow\downarrow\downarrow{}$ (or $uudd$) state and
for Fe on Ru(0001) the $\uparrow\downarrow\uparrow\downarrow{}$-state has been chosen as the compensated configuration.
From the total energy differences shown in Figs.~\ref{fig:tetramers_Rh} and \ref{fig:tetramers_Ru}, it can be seen
that the trend which has been found for the trimers and compact tetramers continues, i.e., the more open a structure is
the larger is the energy gain of the compensated magnetic state. In fact all tetramers considered here prefer a compensated
state. The linear tetramer shows the largest energy gain for compensated states on both Rh(111) and Ru(0001) followed by the
corner-shaped tetramer and the more compact structures which possess similar energy differences. For Fe tetramers on Rh(111)
one can already see the that the $\uparrow\uparrow\downarrow\downarrow{}$-structure is preferred, which is also proposed to be
the magnetic ground state for the full Fe monolayer. In contrast, Fe tetramers on Ru(0001) favor the $\uparrow\downarrow\uparrow\downarrow{}$-state as expected from the antiferromagnetic NN exchange interaction obtained for dimers
(cf.~Fig.~\ref{fig1.2}).

The vertical and lateral structural relaxations which we obtained for these tetramers (not shown) display the same behavior as discussed
for the trimer geometries in section \label{sec:clusters}. From Figs.~\ref{fig:tetramers_Rh} and \ref{fig:tetramers_Ru} it is evident
that the relaxations are critical in order to determine the magnetic ground state. For the considered Fe tetramers on Rh(111) four
out of six configurations experience a sign change of the energy difference upon relaxations, i.e.~the $uudd$ state becomes more
favorable. For the tetramers on Ru(0001) the magnetic configuration even goes from a ferro- to an antiferromagnetic state for all
structures.

\section{Pentamers on R\MakeLowercase{u(0001)}}
\label{sec:pentamers}

\begin{figure}
\includegraphics[width=0.9\linewidth]{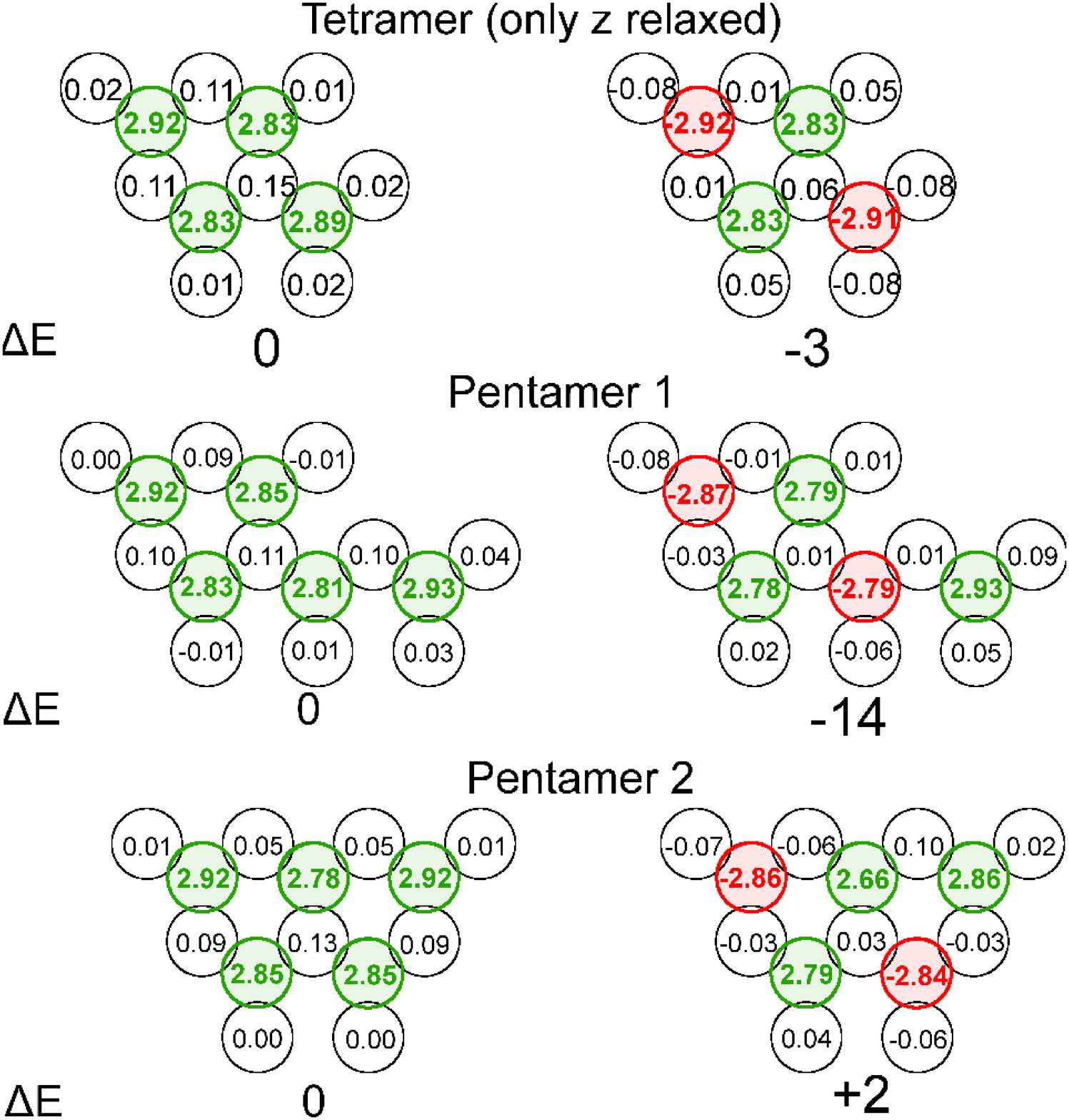}
\caption[fig5]{(color online) (a) Fe tetramer on Ru(0001) in hcp stacking taking only a vertical relaxation into account.
                (b) fully-relaxed Fe pentamers in hcp stacking on Ru(0001). Energies are given in meV/Fe-atom
                with respect to the FM state. Calculations have been performed in the $p(5\times5){}$ and $p(4\times4){}$ unit cell
                for the pentamers and tetramer, respectively.
                }
\label{pents}
\end{figure}

We have seen in section \ref{sec:clusters} that the Fe tetramer on the Ru(0001) surface adopts a ferromagnetic ground
state which was puzzling at first glance since the exchange in the nearest-neighbor Fe dimer is antiferromagnetic.
The explanation is the strengthened direct ferromagnetic interaction between the Fe atoms in the cluster which
dominates over the effect of the Ru substrate. The interaction between the Fe atoms led to reduced lateral
separations within the cluster. In the limit of the complete monolayer, on the other hand, the Fe atoms
are on the ideal two-dimensional lattice sites, the nearest-neighbor exchange is antiferromagnetic, and
a 120$^{o}$ N\'eel state has been proposed as the ground state~\cite{Hardrat}.

In order to see how the interplay of structure, hybridization, and magnetic ground state develops upon increasing
the number of Fe atoms in the cluster we have performed calculations for a pentamer on Ru(0001).
The small energy difference for the $\uparrow\downarrow\downarrow\uparrow{}$- and the FM-state of the tetramer
(cf. Fig. \ref{fig:trimers_tetramers_Ru}) in hcp stacking on Ru(0001) motivates to further
investigate in this direction. Since all compact clusters with ferromagnetic ground states show strongly reduced Fe-Fe distances,
we recalculated the
FM- and $\uparrow\downarrow\downarrow\uparrow{}$- state of the tetramer with a relative vertical relaxation of
$-10{}$ \% but perfect lateral positions as given by the substrate.
As can be seen in Fig.~\ref{pents} with the Fe atoms in this position the $\uparrow\downarrow\downarrow\uparrow{}$-state is slightly
preferred. This can be interpreted as an energy loss for the FM state due to the larger Fe-Fe distances and thus less hybridization.
This finding underlines the crucial importance of structural relaxations in lateral and vertical directions.
One can also enlarge the Fe-Fe spacing by adding a fifth atom to the tetramer.
Indeed for one of the two fully relaxed pentamers on Ru(0001) depicted in Fig.~\ref{pents} the AFM state is
energetically very favorable.
For pentamer 2 the states are nearly degenerate, which can be explained by the frustration of the Fe atoms so that
a N\'eel state might be the ground state. For pentamer 1 the extra Fe atom to the tetramer is not frustrated and
stabilizes the AFM state~\cite{NOTE2}.
This demonstrates how every atom can matter in these clusters for their magnetic ground state.

\section{Infinite atomic chains on R\MakeLowercase{h(111)}}
\label{sec:chains}

\begin{figure*}
\includegraphics[width=0.8\linewidth]{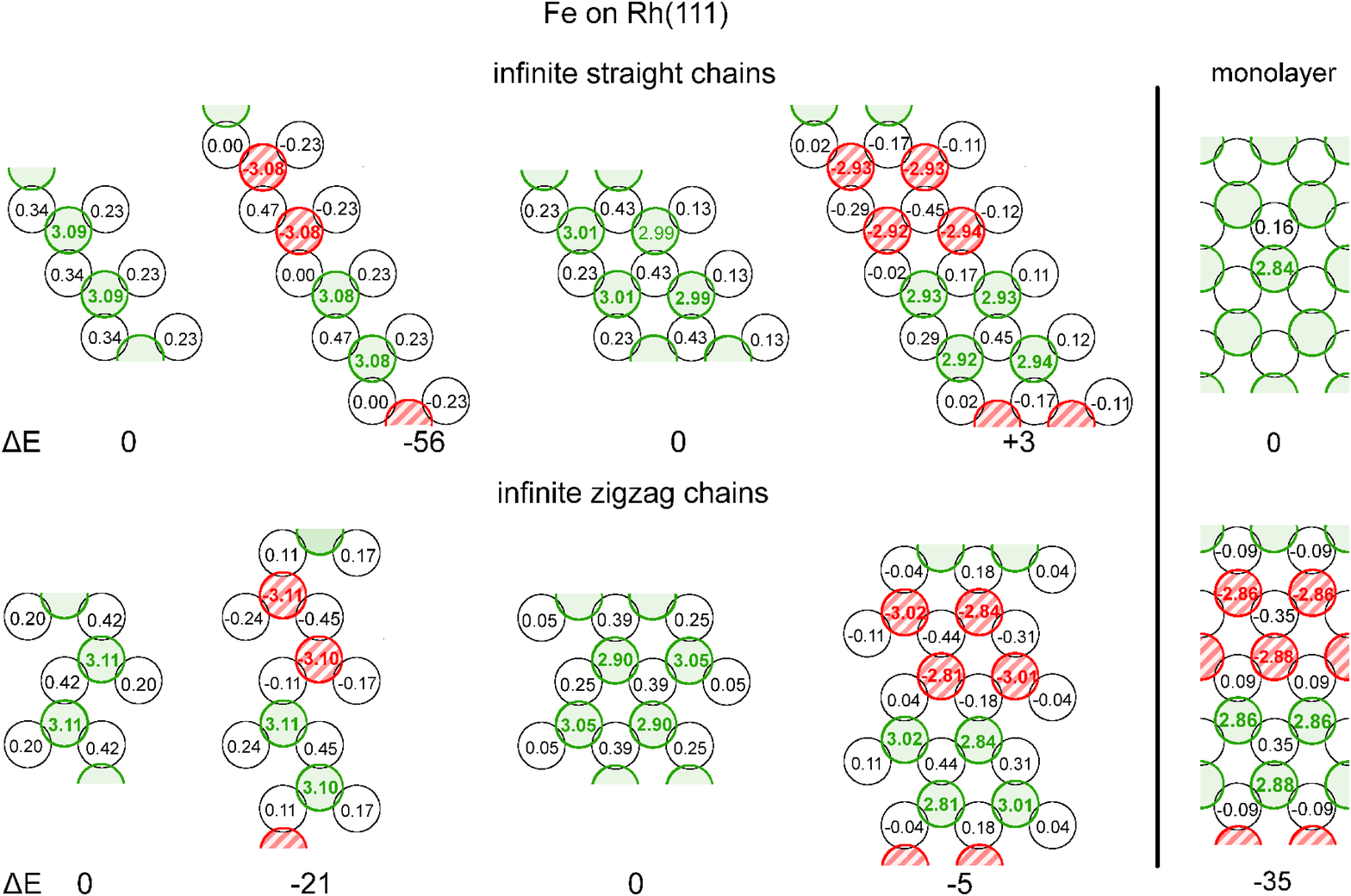}
\caption[fig5]{(color online) Infinite atomic and biatomic Fe chains on Rh(111) in hcp stacking
                along two different crystallographic directions of the surface. The ferromagnetic
                and the $\uparrow\uparrow\downarrow\downarrow$ state have been considered and the energy
                differences are given in meV/Fe-atom.
                Calculations have been performed in the $p(4\times4){}$ and the $c(4\times4){}$ unit cell for
                the zigzag and linear chains, respectively.}
\label{fig6}
\end{figure*}

An interesting question for Fe clusters on the Rh(111) surface concerns the cluster size and structure needed for
the transition to the double-row wise antiferromagnetic state or $uudd$ ($\uparrow\uparrow\downarrow\downarrow$)
state predicted for the full
monolayer~\cite{Hardrat,alzubiPaper}. We explore this issue in the following by considering infinite chains as seen in Fig.~\ref{fig6}.
We find that monoatomic chains favor the $uudd$ state over the FM state, while for biatomic chains, in which every Fe
atom has more nearest neighbors as compared to the monoatomic chains,
both states are nearly degenerate. This behavior is similar to the one found for timers and tetramers, where open structures,
in which Fe atoms have fewer nearest neighbors, tend to compensated states and compact structures to FM states.

\subsection{Monoatomic chains}
We start with monoatomic chains in a straight and a zigzag configuration which can be thought as built from
the linear and corner-shaped trimers considered before.
As one can see both types of monoatomic chains display a clear tendency towards the $\uparrow\uparrow\downarrow\downarrow{}$-state.
The energy difference with respect to the ferromagnetic solution depends strongly on the chain geometry
and it is larger by 35 meV/Fe-atom for the straight chain.
This trend is similar to that observed for the linear and the corner-shaped trimer which showed an
energy gain of 21 and 5 meV/Fe-atom for the antiferromagnetic solutions
, respectively (cf.~Fig.~\ref{fig:trimers_tetramers_Rh}).
The energy difference of 32 meV/Fe-atom in favor of the $\uparrow\uparrow\downarrow\downarrow{}$-state
obtained for a linear tetramer calculated for comparison further strengthens the
conclusion that the exchange coupling in the chains can be understood based on linear clusters.
A look at the density of states for the trimers and the infinite chains (not shown) is in accordance with this
interpretation.
One sees that there is merely a difference in the DOS
for the linear trimer and the straight chain below the Fermi energy and similarly for the zigzag chain and the corner-shaped trimer.
The vertical structural relaxations of the infinite chains and the corresponding trimers are also nearly the same which
explains the similar hybridization with the Rh substrate.

This suggests to apply the exchange constants from the linear and corner-shaped trimer calculations to explain the results of the
infinite chains. Using the nearest and third nearest-neighbor exchange constants $J_1$ and $J_3$
from table \ref{tab3} for the linear chain leads to a total energy difference of
$\Delta E_{uudd-\rm{FM}}=+4J_{1}+8J_{3}=-217{}$ meV, i.e.~$-54{}$ meV/Fe-atom in nearly perfect agreement
with the calculation
(cf.~Fig.~\ref{fig6}). For the zigzag chain an average value of $J_{1}\approx 5{}$ meV is used
because there are two inequivalent nearest-neighbor exchange constants for the corner-shaped trimer.
This results in an energy difference of $\Delta E_{uudd-\rm{FM}}=4J_{1}+8J_{2}=-50{}$ meV, i.e. $-13{}$ meV/Fe atom again in
an appealing agreement with the full calculation.
Compared to the previous delicate behavior of the exchange couplings, this fits remarkably well and is an indication that the reason
for the occurrence of the $uudd{}$ state in the atomic chains is a local effect. So already small clusters with the appropriate shape
such as a linear tetramer may show the $uudd{}$-state.
\subsection{Biatomic chains}
However, for the biatomic Fe chains on Rh(111) the situation turns out to be more complicated due to
the interaction between adjacent strands.
For the straight
biatomic chain the ferromagnetic state is slightly preferred while for the zigzag chain
the $\uparrow\uparrow\downarrow\downarrow{}$ state is more favorable.
However, the energy differences are very small, in particular, with respect to the large energy gain
found for the full monolayer (cf.~Fig.~\ref{fig6}). Within the accuracy of our calculation
the two magnetic states are degenerate for the biatomic chains.
Including a third strand of atoms in our calculation leads to nearly the same energy differences as for
the biatomic chains in both the straight and the zigzag chain structures\footnote{For the three-atomic zigzag-chain one finds $-$5  meV/Fe-atom and for the straight chain
+0.5 meV/Fe-atom}. Apparently, the hybridization with
the additional strands in the chains modifies the exchange coupling considerably as observed in
section \ref{sec:clusters} for the compact clusters. Similar to the comparison of open and compact
clusters we find a tendency towards ferromagnetic states as we move from atomic to biatomic chains.

The lateral relaxations
also play an important role here and emphasize the hybridization between Fe atoms of adjacent strands
in the chain. In particular, in the $\uparrow\uparrow\downarrow\downarrow$ state
Fe atoms with parallel spin alignment relax towards each other which leads
quasi to a formation of tetramers along the chain. The separation between atoms in the two strands is reduced from
the perfect value of 2.70~{\AA} to about 2.45~{\AA} while the spacing along the chain direction amounts
to 2.64~{\AA} between atoms with the same spin and 2.76~{\AA} between atoms with opposite spin.
Therefore, the direct FM exchange between the Fe atoms is strengthened.
In contrast to the compact clusters, however, the ferromagnetic interaction does not prevail which leads to the
very small energy difference to the $\uparrow\uparrow\downarrow\downarrow$ state.
The average vertical relaxation for the biatomic zigzag chain amounts to $-11{}$ \% and $-13{}$ \% in the FM and
in the $\uparrow\uparrow\downarrow\downarrow$ state, respectively.
In order to test the influence of relaxations, we performed a calculation for the biatomic zigzag
chain in which we used the vertical relaxation from the monolayer, i.e. $-6$\% for the FM and $-8$\% for
$\uparrow\uparrow\downarrow\downarrow$ state, and fixed the lateral positions of the Fe atoms according
to the Rh(111) surface. In this structure the $\uparrow\uparrow\downarrow\downarrow$ state becomes favorable
by 16 meV/Fe-atom which demonstrates that the reduced vertical relaxation is crucial in the transition from
the clusters to the full monolayer.

Another important factor for the magnetic ground state is the number of
nearest neighbors for every atom. In the straight biatomic chain both atoms possess four nearest Fe
neighbors while one of the atoms in the zigzag chain interacts with three and the other atom with
five nearest Fe neighbors. In the full monolayer, on the other hand, every Fe atom hybridizes with six
nearest neighbors and the gain of the $uudd$ state is 35 meV/Fe-atom with respect to the ferromagnetic
state.
The vertical relaxation in the full monolayer for the $uudd$ state of $-8$\% is accompanied with a slight
buckling of $0.01$~{\AA}. The atoms also relax in the lateral direction,
i.e~atomic rows with the same spin orientation approach each other by $0.07$~{\AA}
similar to what was observed for the biatomic chains. This breaking of the hexagonal symmetry
of the monolayer should be resolvable in STM experiment using non-magnetic tips. Note that
there is in addition an electronic effect due to the two inequivalent Rh substrate atoms
with different induced magnetic moments (cf.~Fig.~\ref{fig6}) which
should allow the resolution
of the $uudd$ state with conventional STM~\cite{alzubiPaper}.

Overall, we conclude that the transition from small clusters to the full monolayer is quite non-trivial regarding
the magnetic ground state. In particular, structural relaxations play a crucial role which makes it hard to
predict the clusters size at which the $uudd$ state appears. Our calculations hint at the possibility that the
$uudd$ state could already develop for chains with only a few atomic strands and of short length. However, the obtained energy
differences are quite small which makes a definite statement difficult. Therefore, experiments on this system
would be extremely interesting using techniques which are capable of resolving atomic scale
spin structures such as spin-polarized STM or inelastic STS.

\section{Conclusions}

In conclusion, we found a complex trend of the magnetic ground states of Fe clusters on the Rh(111) and the Ru(0001)
surface depending on cluster size, geometry and interatomic distances. Our DFT calculations of Fe dimers demonstrate
that the nearest-neighbor exchange interaction is reduced due to structural relaxation and the strong hybridization
with the substrate and of a similar magnitude as exchange coupling beyond nearest neighbors. On the Rh substrate this
results in a weak ferromagnetic exchange between Fe magnetic moments while it is antiferromagnetic on the Ru surface.
The exchange constants beyond nearest-neighbors display an RKKY-like oscillation and the trend is nearly inverted
for dimers on Rh(111) with respect to those on Ru(0001).

For clusters beyond dimers, there is a competition of
the effect of the substrate and the direct ferromagnetic exchange between Fe atoms in the cluster. Therefore,
the magnetic ground state depends sensitively on the geometry of the cluster and on lateral and vertical
structural relaxations. Small compact
trimers and tetramers become ferromagnetic while open
geometries such as linear and corner-shaped trimers
and tetramers possess antiferromagnetic ground states. This led to the surprising
observation of a ferromagnetic state of compact
trimers and tetramers on Ru(0001) despite the antiferromagnetic NN
exchange in the dimers. Similarly unexpected are the antiferromagnetic states of linear and corner-shaped trimers
on the Rh(111) surface. By mapping the total
energies of the calculations to a Heisenberg model we determined the exchange constants. These depend in a
delicate way on the cluster shape and size due to hybridization within the cluster and with the substrate.
This explains the complex evolution of the magnetic ground state with cluster size.
Adding only a single
atom to the tetramers on Ru(0001) results in the change from a ferromagnetic to an antiferromagnetic state with nearly
compensated magnetic moments.

For Fe clusters on Rh(111) we explored the transition to the predicted $uudd$ magnetic ground state of the full
monolayer by considering infinite chains from one to three strands.
We found that the occurrence of the $uudd$ state
for single atom chains with a large energy gain could be explained
based on the exchange of the open trimers.
However, for Fe chains with two or more strands the interaction
between the strands favors a ferromagnetic state
and the evolution to the $uudd$ state, which is driven by the
interaction with the Rh substrate, is more complicated.
Within our calculations the ferromagnetic and the $uudd$ state are nearly degenerate.

In comparison with previous studies~\cite{Mavro,Ir111,FePd}, our work demonstrates the importance of cluster-substrate hybridization
for the magnetic exchange interaction in clusters on transition-metal surfaces with a partly occupied $d$-band. Therefore,
structural relaxations are crucial to determine the magnetic properties of the clusters. Due to the rich magnetic phase space
of Fe clusters on Rh(111) and Ru(0001) these systems are ideally suited for future experimental studies using spin-polarized
techniques with
high spatial spin resolution and the capability to address spin excitations.

We acknowledge financial support by the Deutsche Forschungsgemeinschaft under project HE3292/8-1.
It is our pleasure to thank Phivos Mavropoulos, Gustav Bihlmayer, and Stefan Bl\"ugel for valuable discussions.

\bibliography{pure_text2}

\begin{thebibliography}{40}
\bibitem[{NOT({\natexlab{a}})}]{NOTE_RKKY}
\eprint{The value for the mixed Fe dimer on Ru(0001) with a spacing of
  3.1~{\AA} has been excluded from the fit since the strong structural
  relaxations (cf.~table~\ref{tab1}) result in exchange constants leading to a
  large deviation from an RKKY-like behavior.}
\bibitem[{NOT({\natexlab{b}})}]{NOTE1}
\eprint{On Rh(111), the energetically most favorable state in hcp (fcc)
  stacking is triangle B, followed by triangle A by 6~meV/Fe-atom
  (33~meV/Fe-atom). The energy difference to the corner-shaped trimer is
  100~meV/Fe-atom (135~meV/Fe-atom) and to the linear one 111~meV/Fe-atom
  (145~meV/Fe-atom). On Ru(0001) one finds the same behaviour with slightly
  different energy differences.}
\bibitem[{NOT({\natexlab{c}})}]{NOTE2}
\eprint{We have also calculated the energy differences for the pentamers using
  a smaller $p(4\times4)$ unit cell and obtained values of $-18$~meV/Fe-atom
  and $-2$~meV/Fe-atom for pentamers 1 and 2, respectively.}
\end{thebibliography}

\end{document}